\documentclass[pra,twocolumn,showpacs,aps]{revtex4-1}
\usepackage{color}
\usepackage{amsmath}
\usepackage{amssymb}
\usepackage{txfonts}
\usepackage{bm}
\usepackage{graphicx} 
\usepackage{enumerate}

\usepackage[unicode,breaklinks]{hyperref}
\hypersetup{
    unicode=true,
    a4paper=true,
    plainpages=false,
    pdftitle={Spinor SPGPE},
    pdfauthor={A. S. Bradley, P. B. Blakie},
    pdfsubject={Spinor SPGPE},
    colorlinks=true,
    linkcolor=blue,
    citecolor=blue,
    filecolor=black,
    urlcolor=blue
}
\newcommand{\ec}{\epsilon_{\rm cut}}

\newcommand{\rr}{{\mathbf r}}
\newcommand{\rrp}{{\mathbf r}^\prime}
\newcommand{\ecut}{\eps^{\rm cut}}
\newcommand{\Hsp}{{\cal H}^{\rm sp}}
\newcommand{\Heff}{{\cal H}^{\rm eff}}
\newcommand{\Veff}{V^{\rm eff}}

\renewcommand{\k}{{\mathbf k}}

\newcommand{\phin}[2]{\phi_{#1}^{#2}}
\newcommand{\phinb}[2]{\bar{\phi}_{#1}^{#2}}

\newcommand{\psin}[2]{\psi_{#1}^{#2}}
\newcommand{\psinb}[2]{\bar{\psi}_{#1}^{#2}}
\newcommand{\lam}{\lambda}
\newcommand{\kap}{\kappa}
\newcommand{\sig}{\sigma}
\newcommand{\gam}{\gamma}
\newcommand{\alp}{\alpha}
\newcommand{\bet}{\beta}
\newcommand{\eps}{\epsilon}

\newcommand{\DP}[2]{\frac{\bar{\delta}#1}{\bar{\delta}#2}}
\newcommand{\DDP}[1]{\frac{\bar{\delta}}{\bar{\delta}#1}}
\newcommand{\DPDP}[2]{\frac{\bar{\delta}^2}{\bar{\delta}#1\bar{\delta}#2}}

\newcommand{\EQ}[1]{\begin{eqnarray}#1\end{eqnarray}}
\newcommand{\mbf}[1]{\mathbf{#1}}

\newcommand{\mint}[1]{\int d^3 #1\;}
\newcommand{\mintp}[2]{\int_{#2} d^3 #1\;}

\newcommand{\QQ}{{\cal Q}}
\newcommand{\PP}{{\cal P}}
\newcommand{\LL}{{\cal L}}

\newcommand{\uu}{{\mathbf u}}
\newcommand{\vv}{{\mathbf v}}

\newcommand{\kk}{\mbf{k}}

\newcommand{\rhoC}{\rho_{C}}

\newcommand{\eref}[1]{(\ref{#1})}
\newcommand{\eeref}[1]{Eq.~(\ref{#1})}
\begin{document}
\title{Stochastic Projected Gross-Pitaevskii equation for spinor and multi-component condensates}
\author{Ashton~S. Bradley and P. Blair Blakie} 
\affiliation{Jack Dodd Center for Quantum Technology, Department of Physics, University of Otago, Dunedin, New Zealand.}
\date{\today}
\begin{abstract}
A stochastic Gross-Pitaevskii equation is derived for partially condensed Bose gas systems subject to binary contact interactions. The theory we present provides a classical-field theory suitable for describing dissipative dynamics and phase transitions of spinor and multi-component Bose gas systems comprised of an arbitrary number of distinct interacting Bose fields. A new class of dissipative processes involving distinguishable particle interchange between coherent and incoherent regions of phase-space is identified. The formalism and its implications are illustrated for two-component mixtures and spin-1 Bose-Einstein condensates. For systems comprised of atoms of equal mass, with thermal reservoirs that are close to equilibrium, the dissipation rates of the theory are reduced to analytical expressions that may be readily evaluated. The unified treatment of binary contact interactions presented here provides a theory with broad relevance for quasi-equilibrium and far-from-equilibrium Bose-Einstein condensates.
\end{abstract}
\maketitle
%\tableofcontents
%============================================================================
\section{Introduction}
Ultracold spinor~\cite{Stenger:1998wt,Lewandowski:2003hs,Schweikhard:2004gc,Saito:2006cm,Liu:2009en,Zhao:2013uk,Endo:2011ip,Kawaguchi:2012bl,StamperKurn:2013ku,Beattie:2013ki} and multi-component~\cite{Matthews1999,Myatt:1997ct,Modugno:2002gz,Simoni:2003hx,Papp:2008fg,Pilch:2009cl,McCarron:2011db,Cho:2013fy}  Bose gases have been  experimentally studied in regimes where an understanding of dissipative and thermal dynamics is necessary. For example: (i) The relaxation dynamics of metastable states arising from the immiscibility of components in a spin-1 sodium condensate  \cite{Miesner:1999vz};
(ii) Energy damping observed in the spin oscillation dynamics of a spinor condensate~\cite{Liu:2009en}, empirically described by an ohmic-like  dissipation term; (iii) Condensation and magnetisation formation dynamics in  a   gas suddenly cooled to below the condensation temperature \cite{Vengalattore:2010ti,Guzman:2011kh}. Many open questions remain about dynamics in this regime \cite{StamperKurn:2013ku}, hampered  by the lack of theoretical techniques to model these systems.
\par
In this paper we develop a theoretical formulation appropriate to the \textit{high-temperature} regime where the condensate exists in the presence of a significant non-condensate population. Our approach is to extend the stochastic projected Gross-Pitaevskii equation (SPGPE)~\cite{Gardiner:2003bk}, which has emerged as a practical quantitative theory of high-temperature single component Bose-Einstein condensates (BECs)~\cite{Bradley:2008gq,Bradley:2005jp,Rooney:2010dp,Rooney:2012gb,Rooney:2014kc}. This extension of the SPGPE theory involves generalising the interactions to arbitrary binary contact interactions and accounting for the multi-component reservoirs that describe the high energy thermalised modes of the system. The resulting spinor/multi-component evolution equations derived are of a Gross-Pitaevskii form, but include additional noise and damping terms to describe the influence of the high energy modes. These equations contain explicit projectors to restrict the dynamical description to the  low energy region of the system, i.e.~the condensate and appreciably occupied low lying modes. An important feature of  multi-component and spinor systems is that new classes of reservoir interactions emerge from the possibility of inter-spin and spin-changing collisions, giving rise to new  types of noise and damping processes.
\par
To put the SPGPE theory into context it is useful to discuss the various theories for describing the thermal dynamics of single component gases, since a variety of methods exist for this case  and a number of comparisons to experiments have been performed. At low to moderate temperatures generalised mean field theories have been developed, and successfully modelled a number of experimental scenarios. The Zaremba-Nikuni-Griffin (ZNG)~\cite{Zaremba:1999iu,Jackson:2001eg,Jackson:2002js,Jackson:2009jo,Jackson:2007gy,Proukakis:2008eo}, projected Gross-Pitaevskii equation (PGPE)~\cite{Davis2001a,Blakie05a,Davis:2006ic,Bezett09a,Bezett09b,Wright:2008ha,Wright:2009eh,Wright:2011ey,Wright:2010pj,Wright:2012ud,Wright:2010dd} (including applications to spinor condensates \cite{Gawryluk:2007ig,Pietila:2010ko}), and number conserving~\cite{Gardiner:2007gj,Billam:2013kb,Mason:2014dd} theories each have advantages for describing BEC evolution, namely, relative ease of handling thermal cloud dynamics, inclusion of many appreciably populated coherent modes, and inclusion of off-diagonal long range order, respectively. At temperatures well below the BEC transition ($T_c$), these effects are essential aspects of finite-temperature BEC physics. However, for temperatures exceeding $\sim T_c/2$, thermal fluctuations from many high-energy incoherent modes become appreciable, motivating an open systems approach. Near the critical point, thermal fluctuations dominate, invalidating the ZNG and number-conserving approaches. Furthermore, while the PGPE approach remains an accurate description of a low-energy coherent fraction near equilibrium, it cannot provide a consistent treatment of the large thermal fraction. In this regime the SPGPE theory is valid and has been shown to provide a quantitative model of experiments (e.g.~see \cite{Weiler:2008eu,Davis:2012hq,Rooney:2013ff}). This theory is valid across the phase transition to Bose-Einstein condensation, enabling studies of critical phenomena such as spontaneous vortex formation~\cite{Weiler:2008eu} (also  see \cite{Blakie:2008is} and \cite{Proukakis:2013dg}). The theory has also been applied to the dissipative dynamics of topological excitations at high temperature~\cite{Rooney:2010dp,Rooney:2011fm,Rooney:2012gb,Garrett:2013gk,Rooney:2013ff}.
A stochastic Gross-Pitaevskii equation formalism has also been developed by Stoof and coworkers~\cite{Stoof:1999tz,Bijlsma:2000vn,Stoof:2001wk,Duine2004,Proukakis06a,Proukakis:2008eo,Cockburn09a,Cockburn10a,Cockburn:2011kw,Cockburn:2011fa,Cockburn:2012gc} that does not impose an explicit projector, but yields a similar description of the low energy region of the system near equilibrium (also see \cite{Damski10a,Su:2013dh,Su:2012jp}).
\par
 The outline of the paper is as follows:
In Sec.~\ref{sys} we outline the class of systems under consideration, and describe the basic decomposition of the interaction Hamiltonian into a $C$-region (coherent), and $I$-region (incoherent). In Sec.~\ref{CRegMast} we derive an equation of motion for the density operator of the $C$-region. In Sec.~\ref{HTFPE} the high-temperature master equation is mapped to a Fokker-Planck equation for the evolution of the Wigner distribution, within the classical-field approximation. In Sec.~\ref{eomSec} the Fokker-Planck equation is mapped to an equivalent stochastic differential equation. The rates of reservoir interaction are evaluated in Appendix \ref{appA}, for the case where the collision involves atoms of equal mass, with $I$-regions described by Bose-Einstein distributions. An explicit treatment of the two-component mixture is given in Sec.~\ref{twoCompMix}, for which all reservoir interaction processes are identified. As a final application, in Sec.~\ref{sec:spin1} we demonstrate the additional energy-damping terms arising in the equations of motion for spin-1 condensates. A concluding discussion of the theory and its implications is given in Sec.~\ref{ConcOut}.

\section{Systems}\label{sys}
In the following, except where stated otherwise, the Einstein summation convention is adopted, whereby repeated greek indices are summed from $-f$ to $f$ for spinor systems of spin $f$, or from 1 to $f$ for mixtures of $f$ components.
\subsection{System Hamiltonian}
The Hamiltonian can be expressed in terms of a set of Bose field operators $\Psi_\sig(\rr)$ (a total of either $2f+1$ or $f$ distinct fields for spinor and mixture systems respectively) with commutation relations
\EQ{
[\Psi_\nu(\rr),\Psi_\sigma^\dag(\rr^\prime)]=\delta_{\nu\sigma}\delta(\rr-\rr^\prime)
}
The spinor Bose gas Hamiltonian 
\EQ{\label{fullHamilS}
H\equiv H_{\rm sp}+H_{\rm int}
}
is written in terms of the interaction Hamiltonian $H_{\rm int}$, and the single-particle Hamiltonian
\EQ{
H_{\rm sp}&=&\mint{\rr}\Psi_\nu^\dag(\rr)\Hsp_\nu\Psi_\nu(\rr),
}
where
\EQ{
\Hsp_\nu \equiv-\frac{\hbar^2\nabla^2}{2m_\nu}+V_\nu(\rr).
}
The possibility of different masses is allowed for here, a feature that has important implications when dealing the multi-component mixtures, such as, for example, the two component system that can be regarded as pseudo-spin 1/2~\cite{Ho:1996kj,Myatt:1997ct}.

In general, the interaction Hamiltonian for the systems we consider can be written as a local two-body interaction  
\EQ{\label{HintDef}
H_{\rm int}=\frac{C^{\lam\nu}_{\kap\sig}}{2}\mint{\rr}\Psi_\lam^\dag(\rr)\Psi^\dag_\nu(\rr)\Psi_\kap(\rr)\Psi_\sig(\rr).
}
\subsubsection{Spinor systems}
For spinor systems multiple internal states of the same atomic species coexist so that $m_\nu\equiv m$, and the potentials incur a linear and quadratic Zeeman shift due to an external bias field that we take to be along $z$, in addition to any spin dependence of the external trapping potentials (although typically optical potentials are used which are spin-independent):
\EQ{
V_\nu(\rr)\equiv V_\nu^{\rm ext}(\rr)-p({\rm f}_z)_{\nu\nu}+q({\rm f}_z)^2_{\nu\nu}.
}
where ${\rm f}_x$, ${\rm f}_y$, and ${\rm f}_z$ are the spin matrices for system with total spin $f$, and $p$ and $q$ parameterise the strength of the linear and quadratic Zeeman shifts respectively.
\par
The interactions are determined by
\EQ{\label{Cdef}
C^{\lam\nu}_{\kap\sig}\equiv\frac{4\pi\hbar^2}{m}\sum_{F=0, 2, \dots,2f}a_{F}\langle f,\lam;f,\nu |{\cal P}_{F}|f,\kappa;f,\sigma\rangle,
}
where 
\EQ{
{\cal P}_{F}\equiv \sum_{J=-{ F}}^{ F}|{ F},J\rangle\langle { F},J|
}
projects onto a two body state with total spin angular momentum $F$, and $a_F$ is $s$-wave scattering length of total spin-$F$ channel~\cite{Kawaguchi:2012bl}. Note that the Clebsch-Gordon coefficients $\langle f,\lam; f,\nu|F,J\rangle$ are only non-zero when $J=\lam+\nu$. The non-vanishing terms in Eqs. \eref{HintDef}, \eref{Cdef} thus conserve the $z$-projection of angular momentum:
\EQ{\label{spinCons}
\lam+\nu=\kap+\sig.
}
Furthermore, the interaction coefficients \eref{Cdef} satisfy the permutation invariance
\EQ{\label{Cperm}
C^{\lam\nu}_{\kap\sig}=C^{\nu\lam}_{\sig\kap}=C^{\sig\kap}_{\nu\lam}=C^{\kap\sig}_{\lam\nu}.
}
\subsubsection{Multi-component mixtures}
In multi-component mixtures distinct atomic species with differing $m_\nu$ are confined by different external potentials $V^{\rm ext}_{\nu}(\rr)$, and interact in the cold-collision regime.
\par
For these systems the interaction matrix elements in \eref{HintDef} are given by (no summation)
\EQ{\label{CMix}
C^{\lam\nu}_{\kap\sig}&\equiv&\frac{\pi\hbar^2 a_{\kap\sig}}{m_{\kap\sig}}(\delta_{\kap\lam}\delta_{\sig\nu}+\delta_{\kap\nu}\delta_{\sig\lam})
}
where $a_{\kap\sig}$ is the $S$-wave scattering length characterising the two-body collision potential for species $\kap$ and $\sig$, and $m_{\kap\sig}^{-1}=m_\kap^{-1}+m_\sig^{-1}$ is the reduced mass. For any parametrisation of the different species in the mixture, the constraints corresponding to \eref{spinCons} become
\EQ{\label{MixCons1}
\lam&=&\kap,\;\;\;\nu=\sig\;\;\;\; \mbox{or}\\
\label{MixCons2}
\lam&=&\sig,\;\;\;\nu=\kap,
}
and the permutation symmetries \eref{Cperm} are immediately evident from the definition \eref{CMix}.
\subsection{Decomposition into $C$- and $I$-regions}
The field operator is decomposed as 
\EQ{\label{psiDec}
\Psi_\lam&=&\phi_\lam+\psi_\lam,
}
where
\EQ{\label{Ppsi}
\phi_\lam&\equiv&{\cal P}_\lam\Psi_\lam,\\
\psi_\lam&\equiv&{\cal Q}_\lam\Psi_\lam,
}
define orthogonal projectors for each spin state, so that ${\cal P}_\lam{\cal Q}_\lam=0$, and the low-energy field $\phi_\lam(\rr)$ is projection onto the set of modes with energies beneath cutoff $\ec^\lam$, defining the $C$-region. 

Expanding $H_{\rm int}$ and utilizing the symmetries of $C^{\lam\nu}_{\sig\kap}$ gives
\begin{subequations}
\label{allprod}
\EQ{\label{prod1}
&&C^{\lam\nu}_{\sig\kap}\Psi^\dag_\lam\Psi^\dag_\nu\Psi_\kap\Psi_\sig=C^{\lam\nu}_{\sig\kap}\Bigg[\phi_\lam^\dag\phi_\nu^\dag\phi_\kap\phi_\sig \\
\label{prod2}
&&+2\phi^\dag_\lam\phi^\dag_\nu\phi_\kap\psi_\sig+2\psi^\dag_\lam\phi^\dag_\nu\phi_\kap\phi_\sig \\
\label{prod3}
&&+\phi^\dag_\lam\phi^\dag_\nu\psi_\kap\psi_\sig+\phi_\lam\phi_\nu\psi^\dag_\kap\psi^\dag_\sig+2\psi^\dag_\lam\phi^\dag_\nu\phi_\kap\psi_\sig+2\psi^\dag_\lam\phi^\dag_\nu\psi_\kap\phi_\sig\;\;\;\;\\
\label{prod4}
&&+2\phi^\dag_\lam\psi^\dag_\nu\psi_\kap\psi_\sig+2\psi^\dag_\lam\psi^\dag_\nu\psi_\kap\phi_\sig\\
\label{prod5}
&&+\psi^\dag_\lam\psi^\dag_\nu\psi_\kap\psi_\sig\Bigg].
}
\end{subequations}
Lines \eref{prod1} and \eref{prod5} can be absorbed in the system Hamiltonians for the $C$ and $I$ regions. The terms that contribute to the reservoir interaction contain either one or two $C$-region operators.  The terms involving three $C$-region operators, \eref{prod2}, do not contribute to collision processes that both conserve energy and momentum, and hence give a vanishing contribution to the reservoir theory.
\par
The Hamiltonian can now be decomposed into $H=H_0+H_I+H_2$ where 
\EQ{
H_0&=&\mint{\rr}\phi^\dag_\lam(\rr)\Hsp_\lam\phi_\lam(\rr)\nonumber\\
&&+\frac{1}{2}\mint{\rr}C^{\lam\nu}_{\kap\sig}\phi_\lam^\dag(\rr)\phi_\nu^\dag(\rr)\phi_\kap(\rr)\phi_\sig(\rr),
}
\EQ{
H_I&=&\mint{\rr}\psi^\dag_\lam(\rr)\Hsp_\lam\psi_\lam(\rr)\nonumber\\
&&+\frac{1}{2}\mint{\rr}C^{\lam\nu}_{\kap\sig}\psi_\lam^\dag(\rr)\psi_\nu^\dag(\rr)\psi_\kap(\rr)\psi_\sig(\rr),
}
and $H_2=H_2^{(1)}+H_2^{(2)}+H_2^{(3)}$ has contributing reservoir interaction terms given by
\EQ{\label{H21def}
H_{2}^{(1)}&=&\mint{\rr}C^{\lam\nu}_{\kap\sig}\Big[\phi^\dag_\lam(\rr)\psi^\dag_\nu(\rr)\psi_\kap(\rr)\psi_\sig(\rr)\nonumber\\
&&+\psi^\dag_\sig(\rr)\psi^\dag_\kap(\rr)\psi_\nu(\rr)\phi_\lam(\rr)\Big]
}
and
\EQ{\label{H22def}
H_{2}^{(2)}&=&\frac{1}{2}\mint{\rr}C^{\lam\nu}_{\kap\sig}\Big[\phi^\dag_\lam(\rr)\phi^\dag_\nu(\rr)\psi_\kap(\rr)\psi_\sig(\rr)\nonumber\\
&&+\phi_\lam(\rr)\phi_\nu(\rr)\psi^\dag_\kap(\rr)\psi^\dag_\sig(\rr)\nonumber\\
&&+2\psi^\dag_\lam(\rr)\phi^\dag_\nu(\rr)\phi_\kap(\rr)\psi_\sig(\rr)\nonumber\\
&&+2\psi^\dag_\lam(\rr)\phi^\dag_\nu(\rr)\psi_\kap(\rr)\phi_\sig(\rr)\Big]
}
A detailed calculation shows that $H_{\rm 2}^{(3)}$ does not lead to collisions that conserve energy and momentum~\cite{Gardiner:2003bk}, and thus these terms are neglected hereafter.
\section{C-region master equation}\label{CRegMast}
The system evolves according to the equation to motion for the density operator
\EQ{
\dot{\rho}&=&-\frac{i}{\hbar}[H_0,\rho]-\frac{i}{\hbar}[H_I,\rho]-\frac{i}{\hbar}[H_2,\rho]\nonumber\\
&\equiv& (\LL_0+\LL_I+\LL_2)\rho.
}
Defining the projection operators for the system density operator
\EQ{
v(t)&=&\PP\rho=\rho_I\otimes {\rm tr}_I(\rho)\equiv\rho_I\otimes\rho_C,\\
w(t)&=&\QQ\rho\equiv (1-\PP)\rho,
}
gives
\EQ{
\dot{v}&=&\PP\left[(\LL_0+\LL_I+\LL_2)(v(t)+w(t))\right],\\
\dot{w}&=&\QQ\left[(\LL_0+\LL_I+\LL_2)(v(t)+w(t))\right].
}
Laplace transforming, and taking the $C$- and $I$-regions as initially uncorrelated ($w(0)=0$) gives
\EQ{
s\tilde{v}(s)-v(0)&=&\PP\left[(\LL_0+\LL_I+\LL_2)(\tilde{v}(s)+\tilde{w}(s))\right],\\
s\tilde{w}(s)&=&\QQ\left[(\LL_0+\LL_I+\LL_2)(\tilde{v}(s)+\tilde{w}(s))\right].
} 
To correctly account for the mean field effect of scattering between $I$- and $C$-region atoms, the new superoperators may be defined
\EQ{
\LL_C\equiv\LL_0+\PP\LL_2\PP,\label{LCdef}\\
\LL_{IC}\equiv\LL_2-\PP\LL_2\PP,\label{LICdef}
}
so that the system evolution $\LL_C$ now includes the effect of \emph{forward scattering}. This procedure retains the mean field contribution to the Hamiltonian evolution of the $C$-region, incurred via scattering from the $I$-region.
In terms of these superoperators, the equations of motion are
\EQ{
s\tilde{v}(s)-v(0)=\PP\left[(\LL_C+\LL_I+\LL_{IC})(\tilde{v}(s)+\tilde{w}(s))\right],\\
\tilde{w}(s)=[s-\QQ(\LL_C+\LL_I+\LL_{IC})]^{-1}\tilde{v}(s).
} 
Inverting via the convolution theorem and making the Markov approximation gives the equation of motion
\EQ{\label{master1}
\dot{v}(t)=\LL_Cv(t)+\left\{\PP\LL_{IC}\int_{-\infty}^0 d\tau\; e^{-(\LL_C+\LL_I)\tau}\QQ\LL_{IC}\right\}v(t),
}
where $\LL_{IC}$ has been neglected in the exponential; the latter approximation is physically justified as the interaction Hamiltonian is a weaker contribution to the evolution over the short time interval included in the integrand, namely, the reservoir correlation time.
\subsection{Hamiltonian terms}
The Hamiltonian terms arise from the $\LL_Cv(t)$ term in \eref{master1} 
\EQ{\label{hamilC}
\dot{\rho}_C\big|_{H_C}&=&-\frac{i}{\hbar}[H_C,\rho_C],
}
where 
\EQ{\label{HCdef}
H_C\equiv H_0+H_F,
}
includes the forward-scattering Hamiltonian
\EQ{\label{HFdef}
H_F\equiv2C^{\lam\nu}_{\lam\nu}\mint{\rr}\langle\psi_\lam^\dag(\rr)\psi_\lam(\rr)\rangle\phi^\dag_\nu(\rr)\phi_\nu(\rr).
}
obtained from \eref{LICdef} and \eref{spinCons}.
It is convenient to include this Hartree-Fock term in an effective potential by defining
\EQ{\label{Veff}
\Veff_\nu(\rr)&\equiv&V_\nu(\rr)+2C^{\lam\nu}_{\lam\nu}\langle\psi_\lam^\dag(\rr)\psi_\lam(\rr)\rangle,
}
and the \emph{effective single particle Hamiltonian}
\EQ{\label{Heffsp}
\Heff_\nu&\equiv&-\frac{\hbar^2\nabla^2}{2m_\nu}+\Veff_\nu(\rr)=\Hsp_\nu+2C^{\lam\nu}_{\lam\nu}\langle\psi_\lam^\dag(\rr)\psi_\lam(\rr)\rangle.
}
\subsection{One-field terms}
The one $C$-field terms in \eref{master1} are due to the one $C$-field terms in \eref{LICdef}, $\LL_{IC}\equiv \LL_{IC}^{(1)}+\LL_{IC}^{(2)}+\LL_{IC}^{(3)}$, namely
\EQ{
\dot{\rho}_C\big|_{(1)}&\equiv&{\rm tr}_I\left[\left\{\PP\LL_{IC}^{(1)}\int_{-\infty}^0 d\tau\; e^{-(\LL_C+\LL_I)\tau}\QQ\LL_{IC}^{(1)}\right\}\rho_C(t)\otimes\rho_I\right].\;\;\;\;\;
}
These terms are usually referred to as the \emph{growth} terms in the $f=0$ theory. In general there is a growth contribution from both one and two field terms, as described below.
To distinguish the two spatio-temporal arguments that arise for field operators in the master equation, use the shorthand
\EQ{
f&\equiv& \hat{f}(\rr,0),\\
\bar{f}&\equiv&\hat{f}(\rr^\prime,\tau)
}
where, as above, the hats are suppressed as the operator character will be clear from the context. The interaction picture field operators are defined as
\EQ{
\phi_\lam(\rr,t)&=&e^{iH_Ct/\hbar}\phi_\lam(\rr,0)e^{-iH_Ct/\hbar},\\
\psi_\lam(\rr,t)&=&e^{iH_It/\hbar}\psi_\lam(\rr,0)e^{-iH_It/\hbar}.
}
Evaluating the ${\cal L}_{IC}^{(1)}$ terms gives
\EQ{\label{masterG1}
\dot{\rho}_C\big|_{(1)}&=&\frac{C^{\lam\nu}_{\kap\sig}C^{\alp\eta}_{\gam\theta}}{\hbar^2}\mint{\rr}\mint{\rr^\prime}\int_{-\infty}^0 d\tau\;\Bigg\{\nonumber\\
&& \langle \psinb{\theta}{\dag}\psinb{\gam}{\dag}\psinb{\eta}{}\psin{\nu}{\dag}\psin{\kap}{}\psin{\sig}{}\rangle [\phi^\dag_\lam,\rho_C\bar{\phi}_\alp] \nonumber\\
&&+\langle \psin{\nu}{\dag}\psin{\kap}{}\psin{\sig}{}\psinb{\theta}{\dag}\psinb{\gam}{\dag}\psinb{\eta}{}\rangle [\phinb{\alp}{}\rhoC,\phin{\lam}{\dag}] \nonumber\\
&&+\langle \psin{\sig}{\dag}\psin{\kap}{\dag}\psin{\nu}{}\psinb{\eta}{\dag}\psinb{\gam}{}\psinb{\theta}{}\rangle [\phinb{\alp}{\dag}\rhoC,\phin{\lam}{}] \nonumber\\
&&+\langle \psinb{\eta}{\dag}\psinb{\gam}{}\psinb{\theta}{}\psin{\sig}{\dag}\psin{\kap}{\dag}\psin{\nu}{}\rangle [\phin{\lam}{},\rho_C\phinb{\alp}{\dag}] \Bigg\}.
}
For thermalised reservoirs with no spin-squeezing, the correlation functions may be Hartree-Fock factorized as, for example,
\EQ{
\langle \psinb{\theta}{\dag}\psinb{\gam}{\dag}\psinb{\eta}{}\psin{\nu}{\dag}\psin{\kap}{}\psin{\sig}{}\rangle&=&\Big[\delta_{\theta\kap}\delta_{\gam\sig}\langle \psinb{\kap}{\dag}\psin{\kap}{}\rangle\langle\psinb{\sig}{\dag}\psin{\sig}{}\rangle\nonumber\\
&&+\delta_{\sig\theta}\delta_{\gam\kap}\langle \psinb{\sig}{\dag}\psin{\sig}{}\rangle\langle\psinb{\kap}{\dag}\psin{\kap}{}\rangle\Big]\langle\psinb{\nu}{}\psin{\nu}{\dag}\rangle\delta_{\eta\nu},\\
\langle \psin{\nu}{\dag}\psin{\kap}{}\psin{\sig}{}\psinb{\theta}{\dag}\psinb{\gam}{\dag}\psinb{\eta}{}\rangle&=&\Big[\delta_{\theta\kap}\delta_{\gam\sig}\langle \psin{\kap}{}\psinb{\kap}{\dag}\rangle\langle\psin{\sig}{}\psinb{\sig}{\dag}\rangle\nonumber\\
&&+\delta_{\sig\theta}\delta_{\gam\kap}\langle \psin{\sig}{}\psinb{\sig}{\dag}\rangle\langle\psin{\kap}{}\psinb{\kap}{\dag}\rangle\Big]\langle\psin{\nu}{}\psinb{\nu}{\dag}\rangle\delta_{\eta\nu}.
}
Hartree-Fock factorizing as above, symmetrising with respect to $C^{\alp\nu}_{\kap\sig}$, and making use of \eeref{spinCons} gives
\EQ{\label{masterG2}
\dot{\rho}_C\big|_{(1)}&=&\frac{\Gamma^{\nu\kap\sig}_{\alp}}{\hbar^2}\mint{\rr}\mint{\rr^\prime}\int_{-\infty}^0 d\tau\;\Bigg\{\nonumber\\
&& \langle \psinb{\nu}{}\psin{\nu}{\dag}\rangle\langle\psinb{\kap}{\dag}\psin{\kap}{}\rangle\langle\psinb{\sig}{\dag}\psin{\sig}{}\rangle [\phi^\dag_\alp,\rho_C\bar{\phi}_\alp] \nonumber\\
&&+\langle \psin{\nu}{\dag}\psinb{\nu}{}\rangle\langle\psin{\kap}{}\psinb{\kap}{\dag}\rangle\langle\psin{\sig}{}\psinb{\sig}{\dag}\rangle [\phinb{\alp}{}\rhoC,\phin{\alp}{\dag}] \nonumber\\
&&+\langle \psin{\nu}{}\psinb{\nu}{\dag}\rangle\langle\psin{\kap}{\dag}\psinb{\kap}{}\rangle\langle\psin{\sig}{\dag}\psinb{\sig}{}\rangle [\phinb{\alp}{\dag}\rhoC,\phin{\alp}{}] \nonumber\\
&&+\langle \psinb{\nu}{\dag}\psin{\nu}{}\rangle\langle\psinb{\kap}{}\psin{\kap}{\dag}\rangle\langle\psinb{\sig}{}\psin{\sig}{\dag}\rangle [\phin{\alp}{},\rho_C\phinb{\alp}{\dag}] \Bigg\},
}
where 
\EQ{\label{GamDef1}
\Gamma^{\nu\kap\sig}_{\alp}&\equiv&\frac{1}{2}(C^{\alp\nu}_{\kap\sig}+C^{\alp\nu}_{\sig\kap})^2,
}
determines the contribution of the particular process in question. 
\par
The one-body Wigner function
\EQ{
W_\sig(\rr,\k)=\mint{\vv}\langle\psin{\sig}{\dag}(\rr+\vv/2)\psin{\sig}{}(\rr-\vv/2)\rangle e^{i\k\cdot\vv}
}
can now be used to evaluate the one-body reservoir correlation functions approximately. In the local density approximation, the thermal equilibrium solution at temperature $T=(\bet k_B)^{-1}$, and spin-dependent chemical potential $\mu_\sig$ is the Bose-Einstein distribution
\EQ{\label{BEdef}
W_\sig(\rr,\k)=[ \exp{\left( \bet[\hbar\omega_\sig(\rr,\k)-\mu_\sig]\right) }-1]^{-1},
}
where the semi-classical energy for \eref{Heffsp} is
\EQ{
\hbar\omega_\sig(\rr,\k)\equiv\frac{\hbar^2\k^2}{2m_\sig}+V_\sig^{\rm eff}(\rr),
}
and the effective potential includes all contributions from external trapping potentials and due to self and cross-interactions through two-body scattering. 
Defining $\uu=(\rr+\rr^\prime)/2$, $\vv=\rr-\rr^\prime$, the two-point correlations can be approximated over the reservoir correlation time-scale of the $\tau$ integration in \eref{masterG2} as
\EQ{\label{stev1}
\langle\psin{\sig}{\dag}(\rr,0)\psin{\sig}{}(\rr^\prime,\tau)\rangle&\approx&\int_{I_\sig} \frac{d^3\kk}{(2\pi)^3}W_\sig(\uu,\k)e^{-i\k\cdot \vv-i\omega_\sig(\uu,\k)\tau},\\ 
\label{stev2}\langle\psin{\sig}{}(\rr,0)\psin{\sig}{\dag}(\rr^\prime,\tau)\rangle&\approx&\int_{I_\sig} \frac{d^3\kk}{(2\pi)^3}[1+W_\sig(\uu,\k)]\nonumber\\
&&\times e^{i\k\cdot \vv+i\omega_\sig(\uu,\k)\tau},\\ 
\label{stev3}\langle\psin{\sig}{}(\rr^\prime,\tau)\psin{\sig}{\dag}(\rr,0)\rangle&\approx&\int_{I_\sig} \frac{d^3\kk}{(2\pi)^3}[1+W_\sig(\uu,\k)]\nonumber\\
&&\times e^{-i\k\cdot \vv-i\omega_\sig(\uu,\k)\tau},\\ 
\label{stev4}\langle\psin{\sig}{\dag}(\rr^\prime,\tau)\psin{\sig}{}(\rr,0)\rangle&\approx&\int_{I_\sig} \frac{d^3\kk}{(2\pi)^3}W_\sig(\uu,\k)e^{i\k\cdot \vv+i\omega_\sig(\uu,\k)\tau},
}
where the integral subscript $I_\sigma$ denotes the semi-classical $I$-region of phase-space for spin component $\sigma$, $I_\sig=\left\{(\rr,\k) | \ecut_\sig\leq \hbar\omega_\sig(\rr,\k)\right\}$.
After making the above approximations for the short time evolution, the time integration may be evaluated using 
\EQ{
\int_{-\infty}^{0}e^{-i\omega\tau}\;d\tau=\pi\delta(\omega)+i{\rm P}(1/\omega)\approx\pi\delta(\omega)
}
where ${\rm P}$ is the principal value integral.
These approximations for the reservoir correlation functions are summarised by defining the growth amplitudes
\EQ{\label{Gp0}
G_{\nu\kap\sig}^{(+)}(\uu,\vv,\eps)&\equiv&\frac{1}{2(2\pi)^8\hbar^2}\mintp{\k_1}{I_\nu}\mintp{\k_2}{I_\kap}\mintp{\k_3}{I_\sig}\nonumber\\
&&\times[1+W_\nu(\uu,\k_1)]W_\kap(\uu,\k_2)W_\sig(\uu,\k_3)e^{i(\k_1-\k_2-\k_3)\cdot\vv}\nonumber\\
&&\times\delta(\eps/\hbar+\omega_\nu(\uu,\k_1)-\omega_\kap(\uu,\k_2)-\omega_\sig(\uu,\k_3)),\nonumber\\
\\\label{Gm0}
G_{\nu\kap\sig}^{(-)}(\uu,\vv,\eps)&\equiv&\frac{1}{2(2\pi)^8\hbar^2}\mintp{\k_1}{I_\nu}\mintp{\k_2}{I_\kap}\mintp{\k_3}{I_\sig}W_\nu(\uu,\k_1)\nonumber\\
&&\times[1+W_\kap(\uu,\k_2)][1+W_\sig(\uu,\k_3)]e^{i(\k_1-\k_2-\k_3)\cdot\vv}\nonumber\\
&&\times\delta(\eps/\hbar+\omega_\nu(\uu,\k_1)-\omega_\kap(\uu,\k_2)-\omega_\sig(\uu,\k_3)),\nonumber\\
}
and the Hamiltonian evolution operator
\EQ{
L_CA\equiv [A,H_C].
}
The growth terms in the master equation now take the form
\EQ{\label{growthMaster}
&&\dot{\rho}_C\big|_{(1)}=\Gamma^{\nu\kap\sig}_{\alp}\mint{\uu}\mint{\vv}\Bigg\{\nonumber\\
&&  \left[\phin{\alp}{\dag}(\uu+\vv/2),\rhoC\left\{G^{(+)}_{\nu\kap\sig}(\uu,\vv,L_C)\phin{\alp}{}(\uu-\vv/2)\right\}\right] \nonumber\\
&&+ \left[\left\{G^{(-)}_{\nu\kap\sig}(\uu,\vv,L_C)\phin{\alp}{}(\uu-\vv/2)\right\}\rhoC,\phin{\alp}{\dag}(\uu+\vv/2)\right] \nonumber\\
&&+\left[\left\{G^{(+)}_{\nu\kap\sig}(\uu,\vv,-L_C)\phin{\alp}{\dag}(\uu-\vv/2)\right\}\rhoC,\phin{\alp}{}(\uu+\vv/2)\right] \nonumber\\
&&+ \left[\phin{\alp}{}(\uu+\vv/2),\rhoC\left\{G^{(-)}_{\nu\kap\sig}(\uu,\vv,-L_C)\phin{\alp}{\dag}(\uu-\vv/2)\right\}\right]\Bigg\},
}
and now have a close formal correspondence with the $f=0$ theory~\cite{Gardiner:2003bk}.
\subsection{Two-field terms}
The two $C$-field terms in \eref{LICdef} give the evolution
\EQ{
\dot{\rho}_C\big|_{(2)}\equiv{\rm tr}_I\left[\left\{\PP\LL_{IC}^{(2)}\int_{-\infty}^0 d\tau\; e^{-(\LL_C+\LL_I)\tau}\QQ\LL_{IC}^{(2)}\right\}\rho_C(t)\otimes\rho_I\right].\;\;\;\;\;\;
}
Evaluating the superoperators, we find
\EQ{\label{masterM1}
&&\dot{\rho}_C\big|_{(2)}=\frac{C^{\lam\nu}_{\kap\sig}C^{\alp\eta}_{\gam\theta}}{\hbar^2}\mint{\rr}\mint{\rr^\prime}\int_{-\infty}^0 d\tau\;\Bigg\{\nonumber\\
&& \langle \psinb{\alp}{\dag}\psinb{\theta}{}\psin{\lam}{\dag}\psin{\sig}{}\rangle [\phi^\dag_\nu\phi_\kap,\rho_C\bar{\phi}_\eta^\dag\bar{\phi}_\gam]
+  \langle \psin{\lam}{\dag}\psin{\sig}{}\psinb{\alp}{\dag}\psinb{\theta}{}\rangle [\bar{\phi}^\dag_\eta\bar{\phi}_\gam\rho_C,\phi_\nu^\dag \phi_\kap]\nonumber\\
&& \langle \psinb{\alp}{\dag}\psinb{\theta}{}\psin{\lam}{\dag}\psin{\kap}{}\rangle [\phi^\dag_\nu\phi_\sig,\rho_C\bar{\phi}_\eta^\dag\bar{\phi}_\gam]
+  \langle \psin{\lam}{\dag}\psin{\kap}{}\psinb{\alp}{\dag}\psinb{\theta}{}\rangle [\bar{\phi}^\dag_\eta\bar{\phi}_\gam\rho_C,\phi_\nu^\dag \phi_\sig]\nonumber\\
&& \langle \psinb{\alp}{\dag}\psinb{\gam}{}\psin{\lam}{\dag}\psin{\sig}{}\rangle [\phi^\dag_\nu\phi_\kap,\rho_C\bar{\phi}_\eta^\dag\bar{\phi}_\theta]
+  \langle \psin{\lam}{\dag}\psin{\sig}{}\psinb{\alp}{\dag}\psinb{\gam}{}\rangle [\bar{\phi}^\dag_\eta\bar{\phi}_\theta\rho_C,\phi_\nu^\dag \phi_\kap]\nonumber\\
&& \langle \psinb{\alp}{\dag}\psinb{\gam}{}\psin{\lam}{\dag}\psin{\kap}{}\rangle [\phi^\dag_\nu\phi_\sig,\rho_C\bar{\phi}_\eta^\dag\bar{\phi}_\theta]
+  \langle \psin{\lam}{\dag}\psin{\kap}{}\psinb{\alp}{\dag}\psinb{\gam}{}\rangle [\bar{\phi}^\dag_\eta\bar{\phi}_\theta\rho_C,\phi_\nu^\dag \phi_\sig]\Bigg\}.\nonumber\\
}
Hartree-Fock factorizing as above, and making use of the symmetries of $C^{\lam\nu}_{\kap\sig}$, we arrive at the master equation
\EQ{
\dot{\rho}_C\big|_{(2)}&=&\frac{{\rm V}_{\sig\nu\gam\eta}^{\lam\kap}}{\hbar^2}\mint{\uu}\mint{\vv}\int_{-\infty}^0 d\tau\;\Bigg\{\nonumber\\
&&\langle\psinb{\kap}{\dag}\psin{\kap}{}\rangle\langle\psinb{\lam}{}\psin{\lam}{\dag}\rangle\left[\phin{\nu}{\dag}\phin{\sig}{},\rhoC\phinb{\eta}{\dag}\phinb{\gam}{}\right]\nonumber\\ \label{Smaster1}
&&+\langle\psin{\kap}{}\psinb{\kap}{\dag}\rangle\langle\psin{\lam}{\dag}\psinb{\lam}{}\rangle\left[\phinb{\eta}{\dag}\phinb{\gam}{}\rhoC,\phin{\nu}{\dag}\phin{\sig}{}\right]\Bigg\},
}
where interaction matrix elements are now defined by (no summation):
\EQ{\label{VDef1}
{\rm V}_{\sig\nu\gam\eta}^{\lam\kap}\equiv (C^{\lam\nu}_{\kap\sig}+C^{\lam\nu}_{\sig\kap})(C_{\kap\eta}^{\lam\gam}+C_{\eta\kap}^{\lam\gam}).
}
This tensor has the permutation symmetries
\EQ{\label{Tsym}
{\rm V}_{\sig\nu\gam\eta}^{\lam\kap}&=&{\rm V}_{\eta\gam\nu\sig}^{\lam\kap}={\rm V}_{\nu\sig\eta\gam}^{\kap\lam}={\rm V}_{\gam\eta\sig\nu}^{\kap\lam},
}
and the spin-conservation condition \eeref{spinCons} now imposes 
\EQ{\label{spinM1}
\lam+\nu&=&\kap+\sig,\\
\label{spinM2}
\lam+\gam&=&\kap+\eta,
}
 or equivalently 
\EQ{\label{spinM3}
\sigma+\gamma&=&\eta+\nu,\\
\label{spinM4}
2(\lam-\kap)&=&\sig+\eta-\nu-\gam,
}
reducing the effective number of indices in the summation to four. The equivalent mass-conservation conditions for mixtures are given by \eref{MixCons1} as
\EQ{\label{MixCons3}
\lam&=&\kap=\gam=\eta,\;\;\;\nu=\sig,\;\;\;\;\mbox{or}\\
\label{MixCons4}
\lam&=&\sig=\gam,\;\;\;\nu=\kap=\eta,\;\;\;\;\mbox{or}\\
\label{MixCons5}
\lam&=&\sig=\eta,\;\;\;\nu=\kap=\gam.
}
 \par
Making use of the approximate short time evolution \eref{stev1}-\eref{stev4}, and evaluating the time integral, the master equation may be written in terms of the \emph{scattering amplitude} defined as
\EQ{\label{Mdef}
M_{\kap\lam}(\uu,\vv,\eps)&\equiv&\frac{1}{2(2\pi)^5\hbar^2}\mintp{\k_1}{I_\kap}\mintp{\k_2}{I_\lam}\nonumber\\
&&\times W_\kap(\uu,\k_1)[1+W_\lam(\uu,\k_2)]e^{i(\k_1-\k_2)\cdot\vv}\nonumber\\
&&\times \delta(\omega_\kap(\uu,\k_1)-\omega_\lam(\uu,\k_2)-\eps/\hbar).
}
The two-field master equation then takes the form
\begin{widetext}
\begin{eqnarray}
\label{fullM}
\dot{\rho}_C\big|_{(2)}={\rm V}_{\sig\nu\gam\eta}^{\lam\kap}\mint{\uu}\mint{\vv}\Bigg\{\left[\phin{\nu}{\dag}(\uu+\vv/2)\phin{\sig}{}(\uu+\vv/2),\rhoC\left\{M_{\kap\lam}(\uu,\vv,L_C)\phin{\eta}{\dag}(\uu-\vv/2)\phin{\gam}{}(\uu-\vv/2)\right\}\right]\nonumber\\
+\left[\left\{M_{\lam\kap}(\uu,\vv,-L_C)\phin{\eta}{\dag}(\uu-\vv/2)\phin{\gam}{}(\uu-\vv/2)\right\}\rhoC,\phin{\nu}{\dag}(\uu+\vv/2)\phin{\sig}{}(\uu+\vv/2)\right]\Bigg\}.
\end{eqnarray}
\end{widetext}
In the scalar theory these terms give rise to the \emph{scattering} master equation~\cite{Gardiner:2003bk}. For colliding indistinguishable particles, the interactions conserve particles in the $C$- and $I$-regions, while transferring energy and momentum between them. When the collision inputs are distinguishable, these terms also allow for particle interchange between $C$- and $I$-regions of distinct spin states (spin-swapping), corresponding to $\lam\neq\kap$ in \eref{spinM4}. The latter process generates an effective growth/loss process for the $C$-field dynamics. A formal decomposition of the two-field interaction into distinct \emph{number} and \emph{energy} damping contributions is given in Sec.~\ref{sec:sgpeDef}.  
 \subsection{Full master equation and equilibrium solution}
 The full master equation is given by \eref{hamilC}, \eref{growthMaster}, \eref{fullM} as
 \EQ{\label{fullMaster}
\dot{\rho}_C=\dot{\rho}_C\big|_{H_C}+\dot{\rho}_C\big|_{(1)}+\dot{\rho}_C\big|_{(2)},
}
and this equation of motion has a grand canonical equilibrium solution.
\subsubsection{Forward-backward relations}
A straightforward calculation shows that the master equation rate functions are related by
\EQ{\label{fbrG}
G^{(-)}_{\nu\kap\sig}(\uu,\vv,\eps)=\exp{\left[\bet(\eps+\mu_\nu-\mu_\kap-\mu_\sig)\right]}G^{(+)}_{\nu\kap\sig}(\uu,\vv,\eps),\;\;\;\;
}
and
\EQ{\label{fbrM}
M_{\kap\lam}(\uu,\vv,\eps)=\exp{\left[-\bet(\eps+\mu_\lam-\mu_\kap)\right]}M_{\lam\kap}(\uu,\vv,-\eps).
}
\subsubsection{Grand canonical equilibrium}
The forward-backward relations give the conditions for the existence of a stationary solution. Without restricting the solution to a global chemical potential, the density operator can be assumed to be a tensor product of grand canonical density operators for each spin state
\EQ{\label{gcsub}
\rho_s\propto \exp{\left[\bet\left(\mu_\sig N_C^\sig-H_C\right)\right]},
}
where 
\EQ{\label{NcDef0}
N_C^\sig&=&\mint{\rr}N_\sig(\rr),
}
and 
\EQ{\label{Nlam}
N_\sig(\rr)=\phi^\dag_\sig(\rr)\phi_\sig(\rr)
}
is the $C$-region number-density operator for spin state $\sig$.
From \eref{fbrG} and \eref{fbrM} one then finds
\begin{widetext}
\EQ{\label{Geq1}
\left\{G^{(-)}_{\nu\kap\sig}(\uu,\vv,L_C)\phin{\alp}{}(\uu-\vv/2)\right\}\rho_s&=&\rho_s\left\{G^{(+)}_{\nu\kap\sig}(\uu,\vv,L_C)\phin{\alp}{}(\uu-\vv/2)\right\}\exp{\left[\bet(\mu_\alp+\mu_\nu-\mu_\kap-\mu_\sig)\right]},\\
\left\{G^{(+)}_{\nu\kap\sig}(\uu,\vv,-L_C)\phin{\alp}{\dag}(\uu-\vv/2)\right\}\rho_s&=&\rho_s\left\{G^{(-)}_{\nu\kap\sig}(\uu,\vv,-L_C)\phin{\alp}{\dag}(\uu-\vv/2)\right\}\exp{\left[-\bet(\mu_\alp+\mu_\nu-\mu_\kap-\mu_\sig)\right]},\\
\label{Meq}
\left\{M_{\lam\kap}(\uu,\vv,-L_C)\phin{\alp}{\dag}(\uu-\vv/2)\phin{\alp}{}(\uu-\vv/2)\right\}\rho_s&=&\rho_s\left\{M_{\kap\lam}(\uu,\vv,L_C)\phin{\alp}{\dag}(\uu-\vv/2)\phin{\alp}{}(\uu-\vv/2)\right\}\exp{\left[\bet(\mu_\lam-\mu_\kap)\right]}.
}
\end{widetext}
Provided the exponential factors reduce to unity, the commutators in the master equation cancel pair-wise and a grand canonical equilibrium solution exists. This yields the growth condition
\EQ{
\mu_\alp+\mu_\nu=\mu_\kap+\mu_\sig
}
for each value of $\alp, \nu, \kap, \sig$, 
and the more restrictive scattering condition
\EQ{
\mu_\lam=\mu_\kap
}
for all $\lam, \kap$. 
The full master equation thus has grand canonical solution when $\mu_\lam\equiv \mu$ for all $\lam$, given by
\EQ{\label{rhoGC}
\rho_s={\cal N}\exp{\left[\bet\left(\mu N_C-H_C\right)\right]},
}
where
\EQ{\label{Ncdef}
N_C=\sum_\sig N^\sig_C,
}
and ${\cal N}$ is a normalisation factor.

\subsection{Mapping to phase-space}
The mapping to a set of equations of motion for classical fields now proceeds along the following lines:
\begin{enumerate}[i)]
\item Linearize the forward-backward relations, appropriate for the high temperature regime.
\item Introduce an appropriate phase-space representation of the density matrix. The Wigner representation provides the correct framework for classical field theory.
\item Define a set of projected functional derivatives that allow mapping of the master equation terms to equivalent terms in a Fokker-Planck equation (FPE) via a set of projected operator correspondences.
\item Carry out the mapping to FPE, neglecting terms of order $(\bet\mu)^2$ or higher as small in the high temperature regime. For positive semidefinite diffusion (as is the case for the FPE obtained in our theory of the spinor system), the FPE can be mapped to a diffusive stochastic process for classical fields variables.
\end{enumerate}
\section{High-temperature Fokker-Planck equation}\label{HTFPE}

\subsection{Wigner representation}
The master equation is mapped to an equation of motion for the Wigner distribution via operator correspondences that are well understood. The formulation for the spinor system is given here, both for completeness, and to establish notation. In the interests of rigour, we reinstate the explicit ``hat" notation for operators, in order to establish the mapping of the quantum theory to classical fields. 
\par
Projected field theory as developed by Gardiner and co-workers~\cite{Gardiner:2003bk,Bradley:2005jp} forms a fundamental framework for the stochastic projected Gross-Pitaevskii theory. The projector first enters dynamics when we consider the Hamiltonian evolution governed by \eref{HCdef}, and now requires explicit definition.
\par
The field operators are expanded in a restricted orthogonal basis as follows:
\EQ{\label{phiExp}
\hat{\phi}_\sig(\rr)\equiv \bar{\sum_n}\hat{a}_{n\sig}Y_{n\sig}(\rr),
}
where the mode operators are bosonic:
\EQ{\label{acom}
[\hat{a}_{n\sig},\hat{a}_{m\nu}^\dag]&=&\delta_{nm}\delta_{\sig\nu},
}
\EQ{\label{acomZero}
[\hat{a}_{n\sig},\hat{a}_{m\nu}]&=&[\hat{a}_{n\sig}^\dag,\hat{a}_{m\nu}^\dag]=0,
}
and the spatial degrees of freedom are expressed in a basis of eigenfunctions of $\Hsp_\sig$:
\EQ{\label{Yeig}
\Hsp_\sig Y_{n\sig}(\rr)=\eps_{n\sig} Y_{n\sig}(\rr).
}
Here the index $n$ is understood to run over all quantum numbers required to specify the eigenmodes. The overbar notation in the sum \eref{phiExp} denotes the restriction to a subset of modes with single-particle energies lying beneath an energy cutoff: $\eps_{n\sig}\leq \ecut_\sig$, as defines the projector. The field operator commutator
\EQ{\label{comm}
[\hat{\phi}_\sig(\rr),\hat{\phi}_\eta^\dag(\rrp)]=\delta_{\sig\eta}\delta_C^\sig(\rr,\rrp)
}
gives a delta function for the $C$-region of spin state $\sig$
\EQ{\label{Pkernel}
\delta_C^\sig(\rr,\rrp)&\equiv&\bar{\sum_{n}}Y_{n\sig}(\rr)Y_{n\sig}^*(\rrp)={\cal P}_\sig(\rr,\rrp),
}
where the latter identifies the kernel of the spatial projection operators via the following definition
\EQ{\label{ProjDef}
{\cal P}_\sig\left\{ h(\rr)\right\}&\equiv&\mint{\rrp}{\cal P}_\sig(\rr,\rrp)h(\rrp).
}
 In practice the basis for each spin state may only differ by the choice of energy cutoff $\ecut_\sig$.
\par
We now introduce the symmetrically ordered quantum characteristic function for the density operator $\hat{\rho}_C$:
\EQ{\label{chiW}
\chi_W[\{\lam_{n\sig},\lam_{n\sig}^*\}]\equiv {\rm tr}\left\{\hat{\rho}_C\exp{\left(\bar{\sum_{m\nu}}\lam_{m\nu}\hat{a}_{m\nu}^\dag-\lam_{m\nu}^*\hat{a}_{m\nu}\right)}\right\},
}
where as usual $\sig$ runs over all distinct fields and $n$ runs over all modes of a given field.
The Wigner function for the system of fields indexed by $\sig$ is then 
\EQ{\label{Wspinor}
W[\{\alpha_{n\sig},\alpha_{n\sig}^*\}]&=&\int \bar{\prod_{m\nu}}\frac{d^2\lam_{m\nu}}{\pi^2}\chi_W[\{\lam_{q\eta},\lam_{q\eta}^*\}]\nonumber\\
&&\times \exp{\left(\bar{\sum_{p\kap}}\lam_{p\kap}^*\alp_{p\kap}-\lam_{p\kap}\alp_{p\kap}^*\right)}.
}
Corresponding to the quantum fields given in \eref{phiExp}, we now define the \emph{classical fields}
\EQ{\label{CfieldDef}
\phi_\sig(\rr)\equiv\bar{\sum_n}\alp_{n\sig}Y_{n\sig}(\rr),
}
where $\alp_{n,\sig}$ are $c$-numbers.
We also require the \emph{projected functional derivatives}
\EQ{\label{PFD1}
\DDP{\phi_\sig(\rr)}&\equiv&\bar{\sum_n}Y_{n\sig}^*(\rr)\frac{\partial}{\partial \alp_{n\sig}},\\
\DDP{\phi_\sig^*(\rr)}&\equiv&\bar{\sum_n}Y_{n\sig}(\rr)\frac{\partial}{\partial \alp_{n\sig}^*}.
}
Using the standard operator correspondences for bosonic modes~\cite{QN}, we obtain the operator correspondences for projected functional calculus of the spinor system
\EQ{\label{ocA}
\hat{\phi}_\sig(\rr)\hat{\rho}&\longleftrightarrow&\left(\phi_\sig(\rr)+\frac{1}{2}\DDP{\phi_\sig^*(\rr)}\right)W,\\
\label{ocB}
\hat{\phi}_\sig^\dag(\rr)\hat{\rho}&\longleftrightarrow&\left(\phi_\sig^*(\rr)-\frac{1}{2}\DDP{\phi_\sig(\rr)}\right)W,\\
\label{ocC}
\hat{\rho}\hat{\phi}_\sig(\rr)&\longleftrightarrow&\left(\phi_\sig(\rr)-\frac{1}{2}\DDP{\phi_\sig^*(\rr)}\right)W,\\
\label{ocD}
\hat{\rho}\hat{\phi}_\sig^\dag(\rr)&\longleftrightarrow&\left(\phi_\sig^*(\rr)+\frac{1}{2}\DDP{\phi_\sig(\rr)}\right)W.
}

\subsection{Hamiltonian evolution}
When mapping the master equation to a phase-space representation, we should expect a close formal correspondence between the Heisenberg equation of motion for the field operator and the classical equation of motion in the Wigner representation, after truncation of third order terms~\cite{Blakie:2008is}.
Evaluating the Heisenberg equation of motion subject to the effective Hamiltonian \eref{HCdef} gives
\EQ{\label{spinorPGPEfield}
i\hbar\frac{\partial \hat{\phi}_\alp(\rr)}{\partial t}&=&\hat{L}_C\hat{\phi}_\alp(\rr)\nonumber\\
&=&{\cal P}_\alp\left\{\Heff_\alp\hat{\phi}_\alp(\rr)+C_{\kap\sig}^{\alp\nu}\hat{\phi}_\nu^\dag(\rr)\hat{\phi}_\kap(\rr)\hat{\phi}_\sig(\rr)\right\},
}
furnishing the PGPE for the $C$-region (including forward scattering) in field operator form. The heuristic approach to obtaining the PGPE involves simply replacing the field operators in this expression with classical fields~\cite{Davis2001a}. Note that for the scalar case this PGPE reduces to the well known projected equation of motion for a single component BEC in the classical limit~\cite{Blakie05a}. 
\par
We now evaluate the Hamiltonian terms via the mappings \eref{ocA}-\eref{ocD}. We also adopt the convention that all quantities arising from the previous sections that now appear \emph{without} hats are the corresponding classical field expressions, for example
\EQ{\label{HCCF}
H_C&=&\mint{\rr}\phi^*_\nu(\rr)\Heff_\nu\phi_\nu(\rr)\nonumber\\
&&+\frac{1}{2}\mint{\rr}C^{\lam\nu}_{\kap\sig}\phi_\lam^*(\rr)\phi_\nu^*(\rr)\phi_\kap(\rr)\phi_\sig(\rr).
}
Making use of the permutation symmetries of $C^{\lam\nu}_{\kap\sig}$, and neglecting the field commutator relative to the particle density, as per the truncated Wigner approximation,
we arrive at
\EQ{\label{HcFPE}
\frac{\partial W}{\partial t}\Bigg|_{H_C}&=&\mint{\rr}\left\{-\DDP{\phi_\alp(\rr)}\left(\frac{-i}{\hbar}\right)L_C\phi_\alp(\rr)+{\rm c.c.}\right\}W,\;\;\;\;
}
where we define the classical field operator $L_C$ as the generator of time evolution via
\EQ{\label{LcNL}
{\cal P}_\alp L_C\phi_\alp(\rr)&\equiv&\DP{H_C}{\phi_\alp^*(\rr)}={\cal P}_\alp\left\{\Heff_\alp\phi_\alp(\rr)+C^{\alp\nu}_{\kap\sig}\phi_\nu^*(\rr)\phi_\kap(\rr)\phi_\sig(\rr)\right\}\nonumber\\
}
corresponding to the quantum operator $\hat{L}_C$. For notational definiteness we do not include the projector in the definition of $L_C$, however, we note that the projector is present in the FPE drift term \eref{HcFPE} due to the action of the projected functional derivative. Our choice of $L_C$ thus means that the projector will be explicitly stated in the $C$-field equations of motion that we derive.
\par
Mapping the drift term in the Fokker-Planck equation, \eref{HcFPE}, to an equivalent stochastic differential equation gives
\EQ{\label{HcEOM}
i\hbar\frac{\partial \phi_\alp(\rr)}{\partial t}&=&{\cal P}_\alp\left\{\Heff_\alp\phi_\alp(\rr)+C^{\alp\nu}_{\kap\sig}\phi_\nu^*(\rr)\phi_\kap(\rr)\phi_\sig(\rr)\right\},
}
namely, the PGPE $C$-field equation corresponding to the Heisenberg equation of motion \eref{spinorPGPEfield}. 
\subsection{One-field terms}
The high temperature treatment requires linearisation of the forward-backward relations. 
The one-field terms are treated by neglecting the $C$-region energy in the forward process, and linearising the forward-backward relation \eref{fbrG} in the reverse process:
\EQ{\label{linGp}
G^{(+)}_{\nu\kap\sig}(\uu,\vv,\eps)&\approx& G^{(+)}_{\nu\kap\sig}(\uu,\vv,0),\\ \label{linGm}
G^{(-)}_{\nu\kap\sig}(\uu,\vv,\eps)&\approx& \left(1-\frac{\mu_{\nu\kap\sig}-\eps}{k_BT}\right)G^{(+)}_{\nu\kap\sig}(\uu,\vv,0).
}
where we introduce the \emph{effective chemical potential}

\EQ{\label{muEff}
\mu_{\nu\kap\sig}&\equiv&\mu_\kap+\mu_\sig-\mu_\nu.
}  
The growth rates \eref{Gp0}, \eref{Gm0} are narrowly peaked functions of $\vv$, allowing integration over this variable. Integrating over \eref{Gp0}, we thus define the rate
\EQ{\label{avG}
G_{\nu\kap\sig}(\uu)&\equiv&\mint{\vv}G_{\nu\kap\sig}^{(+)}(\uu,\vv,0),
}
and make use of the linearised forward-backward relation \eref{linGm}, to find
\begin{widetext}
\EQ{\label{linGrowth}
\frac{\partial\hat{\rho}_C}{\partial t}\Bigg|_{(1)}=\Gamma_\alp^{\nu\kap\sig}\mint{\uu}G_{\nu\kap\sig}(\uu)\Bigg\{\; &&
\left[\left[\hat{\phi}_\alp(\uu),\hat{\rho}_C\right],\hat{\phi}^\dag_\alp(\uu)\right]
+\left[\hat{\phi}_\alp(\uu),\left[\hat{\rho}_C,\hat{\phi}^\dag_\alp(\uu)\right]\right]\nonumber\\
&&-\frac{1}{k_BT}\left(\;\left[\;\left\{(\mu_{\nu\kap\sig}-\hat{L}_C)\hat{\phi}_\alp(\uu)\right\}\hat{\rho}_C,\hat{\phi}^\dag_\alp(\uu)\right]+\left[\;\hat{\phi}_\alp(\uu),\hat{\rho}_C\left\{(\mu_{\nu\kap\sig}+\hat{L}_C)\hat{\phi}^\dag_\alp(\uu)\right\}\right]\right)\;\Bigg\}.
}
Mapping to FPE gives
\EQ{\label{GFPE}
\frac{\partial W}{\partial t}\Bigg|_{(1)}=\mint{\rr}\Gamma_\alp^{\nu\kap\sig}G_{\nu\kap\sig}(\uu)\Bigg\{&&-\DDP{\phi_\alp(\rr)}\left(\frac{(\mu_{\nu\kap\sig}-L_C)\phi_\alp(\rr)}{k_BT}\right)+\DPDP{\phi_\alp(\rr)}{\phi_\alp^*(\rr)}+{\rm c. c.}\Bigg\}W.
}
\end{widetext}
\subsection{Two-field terms}
Although convenient thus far, in its present form, the two-field master equation \eref{fullM} is not suitable for mapping to a high-temperature Fokker-Planck equation as $M_{\lam\kap}$ cannot easily be reduced to a single function to be expanded about $\eps=0$. We can recast the scattering master equation in a form that  symmetrises the rate \eref{Mdef} with respect to $\kap\leftrightarrow\lam$ by replacing ${\rm V}_{\sig\nu\gam\eta}^{\lam\kap}$ with the symmetrized expression
\EQ{\label{LamDef}
\Lambda_{\sig\nu\gam\eta}^{\lam\kap}&\equiv&\frac{1}{2}\left[{\rm V}_{\sig\nu\gam\eta}^{\lam\kap}+{\rm V}_{\nu\sig\eta\gam}^{\kap\lam}\right],
}
so that by construction $\Lambda_{\sig\nu\gam\eta}^{\lam\kap}=\Lambda_{\sig\nu\gam\eta}^{\kap\lam}$,
and we can now define the symmetrised rate
\EQ{\label{calMdef}
M^s_{\lam\kap}(\uu,\vv,\eps)&\equiv&\frac{1}{2}\left[ M_{\lam\kap}(\uu,\vv,\eps)+M_{\kap\lam}(\uu,\vv,\eps) \right],
}
and write the two-field master equation as
\begin{widetext}
\EQ{\label{MasterSpinChange}
\dot{\rho}_C\Big|_{(2)}=\Lambda_{\sig\nu\gam\eta}^{\lam\kap}\mint{\uu}\mint{\vv}\Bigg\{&&\;\;\left[\phin{\nu}{\dag}(\uu+\vv/2)\phin{\sig}{}(\uu+\vv/2),\rhoC\left\{M^s_{\lam\kap}(\uu,\vv,L_C)\phin{\eta}{\dag}(\uu-\vv/2)\phin{\gam}{}(\uu-\vv/2)\right\}\right]\nonumber\\
&&+\left[\left\{M^s_{\lam\kap}(\uu,\vv,-L_C)\phin{\eta}{\dag}(\uu-\vv/2)\phin{\gam}{}(\uu-\vv/2)\right\}\rhoC,\phin{\nu}{\dag}(\uu+\vv/2)\phin{\sig}{}(\uu+\vv/2)\right]\Bigg\}.
}
\end{widetext}
This equation is now in a form suitable for finding a high-temperature master equation as it only depends formally on \emph{one} function, $M^s_{\lam\kap}(\uu,\vv,\eps)$, that may be linearised in the high-temperature regime. Suppressing the $\uu, \vv$ arguments for brevity, the forward-backward relation \eref{fbrM} with \eref{calMdef} now gives
\EQ{\label{fbrMsym}
M^s_{\lam\kap}(\eps)&=&\frac{e^{-\beta\eps}}{2}\Big[e^{\beta(\mu_\lam-\mu_\kap)}M_{\kap\lam}(-\eps)+e^{\beta(\mu_\kap-\mu_\lam)}M_{\lam\kap}(-\eps)\Big]\nonumber\\
&\simeq& e^{-\beta\eps}\Big[M^s_{\lam\kap}(-\eps)\nonumber\\
&&+\frac{\beta(\mu_\lam-\mu_\kap)\left(e^{\beta(\eps+\mu_\kap-\mu_\lam)}M_{\lam\kap}(\eps)-M_{\lam\kap}(-\eps)\right)}{2}\Big],\;\;\;\;\;\;
}
where the high temperature expansion and FBR have been used to simplify the brackets. It is easily shown that the final bracketed term is of order $\bet^2$ as $\eps\rightarrow 0$. Hence, at leading order in $\bet\eps$
\EQ{\label{linMsym}
M^s_{\lam\kap}(\uu,\vv,\eps)&\approx&\left(1-\frac{\eps}{2k_BT}\right)M^s_{\lam\kap}(\uu,\vv,0),
}
namely the standard linearised FBR of the $f=0$ theory~\cite{Gardiner:2003bk},
and the high-temperature master equation for two-field terms is
\begin{widetext}
\EQ{\label{symMasterlin}
\frac{\partial\hat{\rho}_C}{\partial t}\Bigg|_{(2)}=-\Lambda^{\lam\kap}_{\sig\nu\gam\eta}\mint{\uu}\mint{\vv}M^s_{\lam\kap}(\uu,\vv,0)\Bigg\{&&\left[\hat{\phi}_\nu^{\dag}(\uu+\vv/2)\hat{\phi}_\sig(\uu+\vv/2),\left[\hat{\phi}_\eta^{\dag}(\uu-\vv/2)\hat{\phi}_\gam(\uu-\vv/2),\hat{\rho}_C\right]\right]\nonumber\\
&&+\frac{1}{2k_BT}\left[\hat{\phi}_\nu^{\dag}(\uu+\vv/2)\hat{\phi}_\sig(\uu+\vv/2),\left[\left\{\hat{L}_C\hat{\phi}_\eta^{\dag}(\uu-\vv/2)\hat{\phi}_\gam(\uu-\vv/2)\right\},\hat{\rho}_C\right]_+\right]\Bigg\},\;\;\;\;\;
}

where $[\hat{A},\hat{B}]_+\equiv \hat{A}\hat{B}+\hat{B}\hat{A}$ is the anticommutator. 
In the classical field approximation, this maps to the FPE
\EQ{\label{MFPEsym}
\frac{\partial W}{\partial t}\Bigg|_{(2)}=\Lambda^{\lam\kap}_{\sig\nu\gam\eta}\mint{\rr}\mint{\rrp}M^s_{\lam\kap}\left(\frac{\rr+\rrp}{2},\rr-\rrp,0\right)\Bigg[&&-\DDP{\phi_\sig(\rr)}\left(-\frac{\left\{L_C\phi_\gam^*(\rrp)\phi_\eta(\rrp)\right\}}{k_BT}\phi_\nu(\rr)\right)+{\rm c. c.}\nonumber\\
&&+\DDP{\phi_\sig(\rr)}\phi_\nu(\rr)\DDP{\phi^*_\eta(\rrp)}\phi^*_\gam(\rrp)+{\rm c. c.}\nonumber\\
&&-\DDP{\phi_\sig(\rr)}\phi_\nu(\rr)\DDP{\phi_\gam(\rrp)}\phi_\eta(\rrp)+{\rm c.c}\Bigg]W,
}
\end{widetext}
where the action of $L_C$ is defined via \eref{LcNL} to reproduce the algebra of commutators:
\EQ{\label{LCtwoF}
L_C\phi_\nu(\rr)\phi_\sig(\rr)&\equiv& [L_C\phi_\nu(\rr)]\phi_\sig(\rr)+\phi_\nu(\rr)L_C\phi_\sig(\rr),\\
\label{LCtwoF2}
L_C\phi_\nu^*(\rr)&\equiv&-[L_C\phi_\nu(\rr)]^*.
}

\subsection{Canonical two-field interactions}
To make a clearer connection with the single-field theory, we consider the subset of processes for which the $I$-region and $C$-region spin is conserved, $\kap\equiv\lam$, the spin conservation conditions \eref{spinM1}, \eref{spinM2} enforce $\nu=\sig$ and $\gam=\eta$. These processes generate \emph{canonical} two-field interactions involving energy and momentum transfer between $C$- and $I$-regions that independently conserve spin populations. The master equation terms are
\EQ{
\dot{\rho}_C\big|_{(2), can}&=&\frac{{\rm V}_{\sig\eta}^{\lam}}{\hbar^2}\mint{\uu}\mint{\vv}\int_{-\infty}^0 d\tau\;\Bigg\{\nonumber\\
&&\langle\psinb{\lam}{\dag}\psin{\lam}{}\rangle\langle\psinb{\lam}{}\psin{\lam}{\dag}\rangle\left[\phin{\sig}{\dag}\phin{\sig}{},\rhoC\phinb{\eta}{\dag}\phinb{\eta}{}\right]\nonumber\\ \label{Smaster2}
&&+\langle\psin{\lam}{}\psinb{\lam}{\dag}\rangle\langle\psin{\lam}{\dag}\psinb{\lam}{}\rangle\left[\phinb{\eta}{\dag}\phinb{\eta}{}\rhoC,\phin{\sig}{\dag}\phin{\sig}{}\right]\Bigg\},
}
where the three-index tensor elements are defined as
\EQ{\label{VNdiag}
{\rm V}_{\sig\eta}^{\lam}\equiv {\rm V}_{\sig\sig\eta\eta}^{\lam\lam}=(C^{\lam\sig}_{\lam\sig}+C^{\lam\sig}_{\sig\lam})(C_{\lam\eta}^{\lam\eta}+C_{\eta\lam}^{\lam\eta}).
}
The subset of these processes with $\sig\equiv\eta$ correspond to the scenario familiar from the scalar BEC theory~\cite{Gardiner:2003bk}. We can also note the useful symmetry ${\rm V}_{\sig\eta}^{\lam}={\rm V}_{\eta\sig}^{\lam}$.

The scattering master equation for canonical two-field interactions now takes the form
\EQ{\label{MSpinCons}
\dot{\rho}_C&&\big|_{(2),can}=\mint{\uu}\mint{\vv}{\rm V}_{\sig\eta}^{\lam}\Bigg\{\nonumber\\ 
&&\left[N_\sig(\uu+\vv/2),\rhoC\left\{M_{\lam\lam}(\uu,\vv,L_C)N_\eta(\uu-\vv/2)\right\}\right]\nonumber\\
&&+\left[\left\{M_{\lam\lam}(\uu,\vv,-L_C)N_\eta(\uu-\vv/2)\right\}\rhoC,N_\sig(\uu+\vv/2)\right]\Bigg\}.\;\;\;\;\;\;
}
\begin{widetext}
The high temperature master equation is found from the linearized forward-backward relation \eref{linMsym}, and takes the form
\EQ{\label{MMasterlin}
\frac{\partial\hat{\rho}_C}{\partial t}\Bigg|_{(2),can}=-{\rm V}^\lam_{\sig\eta}\mint{\uu}\mint{\vv}M_{\lam\lam}(\uu,\vv,0)\Bigg\{\;&&\left[\hat{N}_\sig(\uu+\vv/2),\left[\hat{N}_\eta(\uu-\vv/2),\hat{\rho}_C\right]\right]\nonumber\\
&&+\frac{1}{2k_BT}\left[\hat{N}_\sig(\uu+\vv/2),\left[\left\{\hat{L}_C\hat{N}_\eta(\uu-\vv/2)\right\},\hat{\rho}_C\right]_+\right]\;\Bigg\}.
}
These terms take the form of a generalised \emph{Quantum Brownian Motion}~\cite{QN}, and act as a source of phase-diffusion~\cite{QN}.
\par
Making use of 
\EQ{\label{LNop}
\hat{L}_C\hat{N}_\sig(\rr)=-i\hbar\nabla\cdot\hat{\mathbf J}_\sig(\rr),
}
where
\EQ{\label{jop}
\hat{\mathbf J}_\sig(\rr)=\frac{i\hbar}{2m_\sig}\left[(\nabla\hat{\phi}^\dag_\sig(\rr))\hat{\phi}_\sig(\rr)-\hat{\phi}_\sig^\dag(\rr)\nabla\hat{\phi}_\sig(\rr)\right],
}
and again neglecting third order terms, the master equation is mapped to the FPE
\EQ{\label{MFPE}
\frac{\partial W}{\partial t}\Bigg|_{(2),can}=\mint{\rr}{\rm V}^\lam_{\sig\eta}M_{\lam\lam}\left(\frac{\rr+\rrp}{2},\rr-\rrp,0\right)\Bigg\{&&-\DDP{\phi_\sig(\rr)}\left(\frac{i\hbar\nabla\cdot\mathbf{J}_\eta(\rrp)}{k_BT}\phi_\sig(\rr)\right)+{\rm c. c.}\nonumber\\
&&+\DDP{\phi_\sig(\rr)}\phi_\sig(\rr)\DDP{\phi^*_\eta(\rrp)}\phi^*_\eta(\rrp)-\DDP{\phi_\sig(\rr)}\phi_\sig(\rr)\DDP{\phi_\eta(\rrp)}\phi_\eta(\rrp)+{\rm c.c}\Bigg\}W.
}
\end{widetext}
\section{Equation of motion}\label{eomSec}
We now carry out the mapping of the high-temperature Fokker-Planck equation given by \eref{HcFPE}, \eref{GFPE}, \eref{MFPEsym} as
\EQ{\label{FullFPE}
\frac{\partial W}{\partial t}&=&\frac{\partial W}{\partial t}\Bigg|_{H_C}+\frac{\partial W}{\partial t}\Bigg|_{(1)}+\frac{\partial W}{\partial t}\Bigg|_{(2)},
}
 to an equivalent stochastic differential equation.
\subsection{Stochastic projected Gross-Pitaevskii equation}\label{sec:sgpeDef}
Carrying out the mapping to an equivalent stochastic diffusion process~\cite{Gardiner:2009wp}, the equation of motion in Stratonovich form can now be obtained as
\begin{widetext}
\begin{subequations}
\label{alleomS}
\EQ{\label{eomSa}
(S)d\phi_\alp(\rr)&=&-\frac{i}{\hbar}{\cal P}_\alp\left\{ L_C\phi_\alp(\rr)\right\}dt\\
\label{eomSb}
&&+{\cal P}_\alp\left\{\frac{\Gamma_\alp^{\nu\kap\sig}G_{\nu\kap\sig}(\rr)}{k_BT}(\mu_{\nu\kap\sig}-L_C)\phi_\alp(\rr)dt+dW_{\alp}(\rr,t)\right\}\\
\label{eomSc}
&&+{\cal P}_\alp\Bigg\{\mint{\rrp}\Lambda^{\lam\kap}_{\alp\nu\gam\eta}M^s_{\lam\kap}\left(\frac{\rr+\rrp}{2},\rr-\rrp,0\right)\left(-\frac{\left\{L_C\phi_\gam^*(\rrp)\phi_\eta(\rrp)\right\}}{k_BT}\phi_\nu(\rr)\right)dt+i\phi_\nu(\rr)dU_\alp^\nu(\rr,t)\Bigg\}.
}
\end{subequations}
\end{widetext}
Here the chemical potentials are defined in \eref{muEff}, and the nonlinear operator $L_C$ is defined by \eref{LcNL}, \eref{HCCF}, and \eref{LCtwoF}, \eref{LCtwoF2}. The one-field rates are given by \eref{GamDef1}, \eref{avG}, and the two-field rates are given by \eref{LamDef}, \eref{VDef1}, and \eref{calMdef}, \eref{Mdef}.

The noises $dW_\alp$, and $dU_\alp^\nu$ are independent Gaussian Wiener processes with non-vanishing correlation functions 
\EQ{\label{dW}
dW^*_\alp(\rr,t)dW_\eta(\rrp,t)=2\Gamma_\alp^{\nu\kap\sig}G_{\nu\kap\sig}(\rr)\delta_C^\alp(\rr,\rrp)\delta_{\alp\eta}dt,
}
\EQ{\label{dU}
dU_\alp^\nu(\rr,t)dU_\gam^\eta(\rrp,t)=2\Lambda^{\lam\kap}_{\alp\nu\gam\eta}M^s_{\lam\kap} \left(\frac{\rr+\rrp}{2},\rr-\rrp,0\right)dt.\;\;\;\;\;\;\;
}
%%--------------------------------------------------------------------
\begin{figure}[!t]{
\begin{center} 
\includegraphics[width=\columnwidth]{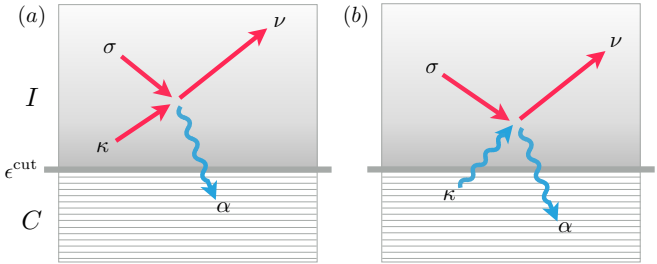}
\caption{Schematic of reservoir-interaction processes for the binary contact interaction Hamiltonian. Collision processes described by (a) the one-field terms, and (b) the two-field terms given by \eref{eomSb} and \eref{eomSc} respectively. In general the latter processes include  grand-canonical interactions with the reservoir involving both number and energy exchange between $C$ and $I$.
\label{fig1}}
\end{center}}
\end{figure}
%%---------------------------------------------------------------
Some comments are in order. 
\begin{enumerate}[(i)]
\item The term \eref{eomSa} gives the PGPE \eref{HcEOM} for the $C$-field evolution according to Hamiltonian \eref{HCCF}. In addition to the $C$-field evolution, this equation of motion allows for spatial and temporal evolution of the $I$-regions through the effective potential \eref{HFdef}. These terms are not treated further in the present work.
\item \emph{Simple growth SPGPE:} If the reservoirs may be well approximated by Bose-Einstein distributions, and the two-field interactions may be neglected --- i.e. the $C$-regions are not too far from equilibrium --- then we recover a description known as the simple growth SPGPE~\cite{Bradley:2008gq}, involving only the terms \eref{eomSa} and \eref{eomSb} and noise \eref{dW}.
\item \emph{Decomposition of two-field interactions:} The two-field terms, \eref{eomSc} and noise \eref{dU}, may be decomposed using \eref{VNdiag} as
\EQ{\label{LamDecomp}
\Lambda^{\lam\kap}_{\alp\nu\gam\eta}&=&\bar{\Lambda}^{\lam\kap}_{\alp\nu\gam\eta}+{\rm V}_{\alp\eta}^{\lam}\delta_{\lam\kap}\delta_{\alp\nu}\delta_{\gam\eta},
}
where the first term
\EQ{\label{LamBar}
\bar{\Lambda}^{\lam\kap}_{\alp\nu\gam\eta}&\equiv&\Lambda^{\lam\kap}_{\alp\nu\gam\eta}-{\rm V}_{\alp\eta}^{\lam}\delta_{\lam\kap}\delta_{\alp\nu}\delta_{\gam\eta},
}
involves number and energy dissipation (grand canonical), and the second term is purely an energy-dissipative process (canonical). The latter class of processes include the energy transfer interaction known from the scalar SPGPE theory~\cite{Gardiner:2003bk}, and additional interactions involving collisions between \emph{distinguishable} particles.
\end{enumerate}
In Figure \ref{fig1} we give a schematic summary of the processes involved in reservoir interactions for spinor and multi-component systems. 

\subsection{SPGPE for spinors and equal-mass mixtures with quasi-static reservoirs}
For components with equal mass in the quasi-static reservoir regime, the rates can be evaluated analytically, as given in Appendix \ref{appA}. This treatment of the rate functions will also provide a good approximation for mixtures with small mass imbalance. Under these assumptions, and making use of the decomposition \eref{LamDecomp}, the full SPGPE \eref{alleomS} can be reduced to the SPGPE
\begin{widetext}
\begin{subequations}
\label{alleomS2}
\EQ{\label{eomS2a}
(S)d\phi_\alp(\rr)&=&-\frac{i}{\hbar}{\cal P}_\alp\left\{ L_C\phi_\alp(\rr)\right\}dt \\
\label{eomS2b}
&&+{\cal P}_\alp\left\{\frac{\Gamma_\alp^{\nu\kap\sig}G_{\nu\kap\sig}}{k_BT}(\mu_{\nu\kap\sig}-L_C)\phi_\alp(\rr)dt+dW_{\alp}(\rr,t)\right\} \\
\label{eomS2c}
&&+{\cal P}_\alp\Bigg\{\mint{\rrp}{\rm V}^{\lam}_{\alp\eta}M_{\lam\lam}\left(\rr-\rrp\right)\left(\frac{i\hbar\nabla\cdot \mathbf{J}_\eta(\rrp)}{k_BT}\phi_\alp(\rr)\right)dt
+i\phi_\alp(\rr)dU_\alp(\rr,t)\Bigg\} \\
\label{eomS2d}
&&+{\cal P}_\alp\Bigg\{\mint{\rrp}\bar{\Lambda}^{\lam\kap}_{\alp\nu\gam\eta}M^s_{\lam\kap}\left(\rr-\rrp\right)\left(-\frac{\left\{L_C\phi_\gam^*(\rrp)\phi_\eta(\rrp)\right\}}{k_BT}\phi_\nu(\rr)\right)dt
+i\phi_\nu(\rr)dU_\alp^\nu(\rr,t)\Bigg\},
}
\end{subequations}
where the noise correlations are 
\EQ{\label{dWqs}
dW^*_\alp(\rr,t)dW_\eta(\rrp,t)=2\Gamma_\alp^{\nu\kap\sig}G_{\nu\kap\sig}\delta_C^\alp(\rr,\rrp)\delta_{\alp\eta}dt,
}
\EQ{\label{dVqs}
dU_\alp(\rr,t)dU_\eta(\rrp,t)=2{\rm V}^\lam_{\alp\eta}M_{\lam\lam} \left(\rr-\rrp\right)dt,
}
\EQ{\label{dUqs}
dU_\alp^\nu(\rr,t)dU_\gam^\eta(\rrp,t)=2\bar{\Lambda}^{\lam\kap}_{\alp\nu\gam\eta}M^s_{\lam\kap} \left(\rr-\rrp\right)dt.
}
The rate $G_{\nu\kap\sig}$ derived in Appendix \ref{appA} is independent of $\rr$,
\EQ{\label{Gfinal}
G_{\nu\kap\sig}&=&G_0\bar{G}_{\nu\kap\sig}\equiv G_0 e^{\beta(\mu_\sig-\ecut_\sig)+\beta(\mu_\kap-\ecut_\kap)}\sum_{r=0}^\infty e^{\beta r(\mu_\nu-\ecut_\sig-\ecut_\kap)}\Phi\left[e^{\beta(\mu_\sig-\ecut_\sig)},1,r+1\right]\Phi\left[e^{\beta(\mu_\kap-\ecut_\kap)},1,r+1\right],
}
where the magnitude is
\EQ{\label{G0}
G_0=\frac{m^3}{(2\pi)^3\hbar^7\beta^2},
}
and where 
$\Phi[z,x,a]=\sum_{k=0}^\infty z^k/(a+k)^x$ is the \emph{Lerch transcendent}. The two-field rate $M^s_{\lam\kap}((\rr+\rrp)/2,\rr-\rrp,0)$ has been reduced to the non-local form
\EQ{\label{Mfinal}
M^{s}_{\kap\lam}(\rr)\equiv\frac{{\cal N}_0\bar{{\cal N}}^s_{\kap\lam}}{(2\pi)^3}\mint{\kk}\frac{e^{i\kk\cdot\rr}}{|\kk|},
}
where 
\EQ{\label{Ndef}
{\bar {\cal N}}^s_{\kap\lam}&\equiv&\frac{1}{2}\sum_{p=1}^\infty \left(e^{p\beta(\mu_\kap-\bar{\epsilon}_{\kap\lam})}\Phi\left[e^{\beta(\mu_\lam-\bar{\epsilon}_{\kap\lam})},1,p \right]+e^{p\beta(\mu_\lam-\bar{\epsilon}_{\kap\lam})}\Phi\left[e^{\beta(\mu_\kap-\bar{\epsilon}_{\kap\lam})},1,p \right]\right),\\
\label{NoDef}
{\cal N}_0&=&\frac{\pi m^2}{(2\pi)^2\hbar^5\beta},
}
and the cutoff $\bar{\epsilon}_{\kap\lam}$ is given by \eref{ecutBar}.
\end{widetext}

Equations \eref{alleomS2}, \eref{dWqs}, \eref{dVqs}, \eref{dUqs} give a generalization of the scalar SPGPE theory~\cite{Gardiner:2003bk}, as implemented in \cite{Bradley:2008gq} and \cite{Rooney:2012gb}, in a form suitable for numerical simulations of dissipative dynamics and quench phenomena in spinor and multi-component systems. We now apply this formalism to the simplest examples, namely the two-component mixture, and the spin-1 system.
\subsection{Two-component mixtures}\label{twoCompMix}
The simplest example involves a two-component mixture, with equal (or nearly equal) constituent masses. Such a setup occurs for different hyperfine levels of $^{87}$Rb, and has been studied at length~\cite{Ho:1996kj,Hall:1998vx,Papp:2008fg,Sabbatini:2011gu}. The $C$-field region of the system is described by Bose fields $\phi_j(\rr)$ for $j=1,2$.
\par
\emph{Hamiltonian terms}.--- Evaluating \eref{eomS2a}, for $m_{12}=m/2$, we find
\EQ{\label{Hmix}
\hbar d\phi_j(\rr)\Big|_{H_C}&=&-i{\cal P}_j \{L_C\phi_j(\rr)\}dt,
}
where
\EQ{\label{Lcmix}
L_C\phi_j(\rr)&\equiv&\Heff_j\phi_j(\rr)+\frac{4\pi\hbar^2a_{jj}}{m}|\phi_j|^2\phi_j\nonumber\\
&&\;\;\;\;\;\;\;+\frac{4\pi\hbar^2a_{j(3-j)}}{m}|\phi_{3-j}|^2\phi_{j},
}
giving the PGPE for Hamiltonian $C$-field evolution~\cite{Blakie:2008is}.
\par
\emph{One-field dissipation}.--- Evaluating the dimensionless rates \eref{Gfinal}, and making use of \eref{spinCons}, and the thermal de-Broglie wavelength $\lam_{dB}\equiv\sqrt{2\pi\hbar^2/mk_BT}$, \eref{eomS2b} gives
\EQ{\label{oneFieldTerms}
\hbar d\phi_j(\rr)\Big|_{(1)}&=&{\cal P}_j\Big\{\gamma_{j}(\mu_j-L_C)\phi_j(\rr)dt+\hbar dW_j(\rr,t)\Big\},\;\;\;\;\;
}
where
\EQ{\label{mixOneF}
\gam_{j}&=&\frac{1}{\pi\lam_{dB}^2}\left[\sig_{jj}\bar{G}_{jjj}+\sig_{j(3-j)}\bar{G}_{j(3-j)j}\right],
}
and where $\sig_{jk}=4\pi a_{jk}^2(1+\delta_{jk})$ is the total cross-section for scattering between $j$ and $k$. The noise correlation is 
\EQ{\label{mixNoise}
dW_j^*(\rr,t)dW_k(\rr,t)&=&\frac{2\gamma_{j}k_BT}{\hbar}\delta_C^j(\rr,\rrp)\delta_{jk}dt.\;\;\;\;
}
The first term in \eref{mixOneF} recovers the known expression for the damping rate of a scalar BEC~\cite{Bradley:2008gq}, and the second term accounts for the different scattering cross-section of distinguishable particles [See Figure \ref{fig2} (a)].  
\par
\emph{Two-field dissipation: energy damping}.--- 
We first consider the canonical energy dissipation terms, given by \eref{eomS2c}. Evaluating the rates, we find
\EQ{\label{2mixED}
(S)\hbar d\phi_j(\rr)\Big|_{(2), {\rm V}}&=&{\cal P}_j\Big\{-iV_j^M(\rr)\phi_j(\rr)dt\nonumber\\
&&+i\hbar \phi_j(\rr)dU_j(\rr,t)\Big\},
}
where the energy-damping potential is
\EQ{\label{VMj2}
V_j^M(\rr)&=&-\hbar\mint{\rrp} {\cal M}_{j}(\rr-\rrp)\nabla\cdot \mathbf{J}_j(\rrp),
}
with 
\EQ{\label{MjDef}
{\cal M}_{j}(\rr)&=&\left(2\sig_{jj}\bar{\cal N}_{jj}+\sig_{j(3-j)}\bar{\cal N}_{(3-j)(3-j)}\right)\int \frac{d^3\kk}{(2\pi)^3}\frac{e^{i\kk\cdot\rr}}{|\kk|}.\;\;\;\;\;\;
}
The noise correlation is
\EQ{\label{dUjDef}
dU_j(\rr,t)dU_k(\rrp,t)&=&\frac{2 k_BT}{\hbar}{\cal M}_j(\rr-\rrp)\delta_{jk}dt.
}
Using \eref{LerchId} we have that $\bar{\cal N}_{jj}=[e^{\beta (\ecut_j-\mu_j)}-1]^{-1}$, and the first term in \eref{MjDef} recovers the scalar BEC result. [See Figure \ref{fig2} (b)]
\par
\emph{Two-field dissipation: particle exchange terms}.--- This process involves distinct particles swapping between respective $C$- and $I$-regions, given by \eref{eomS2d}. Consequently this process involves both energy and number exchange, as do the one-field terms. 
\par
The mass-conservation rules \eref{MixCons3}-\eref{MixCons5} show the contributing interactions involving particle swapping for $d\phi_\alp$ are $\Lambda^{\alp\lam }_{\alp\lam\alp\lam}$ and $\Lambda^{\alp\lam}_{\alp\lam\lam\alp}$. Due to the form of $C^{\lam\nu}_{\kap\sig}$ in \eref{CMix} we also have $\Lambda^{\alp\lam}_{\alp\lam\lam\alp}=\Lambda^{\alp\lam }_{\alp\lam\alp\lam}$, so that the net term involves $L_C\phi^*_\gam\phi_\eta+L_C\phi^*_\eta\phi_\gam$. Evaluating the term, we find
\EQ{\label{LC12}
-L_C\phi_2^*(\rr)\phi_1(\rr)-L_C\phi_1^*&&(\rr)\phi_2(\rr)=i\hbar\nabla\cdot( \mathbf{J}_{12}(\rr)+\mathbf{J}^*_{12}(\rr))\nonumber\\
&&+(\phi_2^*(\rr)\phi_1(\rr)-\phi_2(\rr)\phi_1^*(\rr))V_{12}(\rr),\;\;\;\;\;\;
}
where
\EQ{\label{j12Def}
\mathbf{J}_{12}(\rr)&=&\frac{i\hbar}{2m}\left\{[\nabla\phi_2^*(\rr)]\phi_1(\rr)-\phi_2^*(\rr)\nabla\phi_1(\rr)\right\},\\
\label{V12Def}
V_{12}(\rr)&=&\Veff_2(\rr)-\Veff_1(\rr)\nonumber\\
&&+(4\pi\hbar^2/m)(a_{22}-a_{12})|\phi_2(\rr)|^2\nonumber\\
&&+(4\pi\hbar^2/m)(a_{12}-a_{22})|\phi_1(\rr)|^2,
}
describe kinetic and potential energy contributions respectively. Note that both terms in \eref{LC12} are explicitly imaginary, and the term takes the form of a potential. However, as the potential also acts as a coupling between the two fields, particle transfer can occur.
We then find, in general
\EQ{\label{2FND}
(S)\hbar d\phi_j(\rr)\Big|_{(2), \bar{\Lambda}}&=&{\cal P}_j\Big\{-iV_{j(3-j)}^M(\rr)\phi_{3-j}(\rr)dt\nonumber\\
&&+i\hbar \phi_{3-j}(\rr)dU_j^{3-j}(\rr,t)\Big\},
}
where
\EQ{\label{V12}
V_{j(3-j)}^M(\rr)&=&-\hbar\mint{\rrp}{\cal M}_{j(3-j)}(\rr-\rrp)\nonumber\\
&&\times\Big[\nabla\cdot \mathbf{J}_{j(3-j)}^R(\rrp)\nonumber\\
&&+\frac{1}{2i\hbar}(\phi_{3-j}^*(\rrp)\phi_j(\rrp)\nonumber\\
&&-\phi_{3-j}(\rrp)\phi_j^*(\rrp))V_{j(3-j)}(\rrp)\Big],\;\;\;\;\;\;\;\;
}
with
\EQ{\label{MjkDef}
{\cal M}_{j(3-j)}(\rr)&=&2\sig_{j(3-j)}\bar{\cal N}^s_{j(3-j)}\int \frac{d^3\kk}{(2\pi)^3}\frac{e^{i\kk\cdot\rr}}{|\kk|},\;\;\;\;
}
and $\mathbf{J}_{j(3-j)}^R(\rr)=(\mathbf{J}_{j(3-j)}(\rr)+\mathbf{J}_{j(3-j)}^*(\rr))/2$.
The noise correlation is
\EQ{\label{dUjDef2}
dU_j^{3-j}(\rr,t)dU_k^{3-k}(\rrp,t)&=&\frac{2 k_BT}{\hbar}{\cal M}_{j(3-j)}(\rr-\rrp)\delta_{jk}dt.\;\;\;\;
}
Using the symmetry $V_{12}^M(\rr)=V_{21}^M(\rr)$  and considering the evolution of the density difference due only to the potential in \eref{2FND}, we find
\EQ{
\frac{\partial}{\partial t} [N_1(\rr)-N_2(\rr)]&=&\frac{2iV_{12}^M(\rr)}{\hbar}[\phi_2^*(\rr)\phi_1(\rr)-\phi_1^*(\rr)\phi_2(\rr)],\nonumber\\\label{N12}
}
and it is clear that these terms cause particle exchange between the two $C$-regions, mediated by the $I$-region interaction [See Figure \ref{fig2} (c)].
\subsection{Spin-1 Bose-Einstein condensates}\label{sec:spin1}
For spin-1, the system is described by the spinor wave function  $\mathbf{\phi}(\rr)\equiv [\phi_{1}(\rr), \phi_0(\rr), \phi_{-1}(\rr)]^T$, and the  matrix elements of the interaction Hamiltonian can be written as~\cite{Kawaguchi:2012bl}
\EQ{\label{s1Hint}
C^{\lam\nu}_{\kap\sig}=c_0\delta_{\lam\kap}\delta_{\nu\sig}+c_1\sum_{j=x,y,z}({\rm  f}_j)_{\lam\kap}({\rm f}_j)_{\nu\sig},
}
where the spin matrices are
\begin{widetext}
\EQ{{\rm f}_x=
\frac{1}{\sqrt{2}}\left[ \begin{array}{ccc}
0 & 1 & 0 \\
1 & 0 & 1 \\
0 & 1 & 0 \end{array} \right];
\;\hspace{0.2cm}
{\rm f}_y=
\frac{i}{\sqrt{2}}\left[ \begin{array}{ccc}
0 & -1& 0 \\
1 & 0 & -1 \\
0 & 1 & 0 \end{array} \right];
\;\hspace{0.2cm}
{\rm f}_z=
\left[ \begin{array}{ccc}
1 & 0 & 0 \\
0 & 0 & 0 \\
0 & 0 & -1 \end{array} \right],
}
\end{widetext}
and the interaction parameters are
\EQ{\label{S1int}
c_0&=&\frac{g_0+2g_2}{3}=\frac{4\pi\hbar^2}{3m}(a_0+2a_2),\\
c_1&=&\frac{g_2-g_0}{3}=\frac{4\pi\hbar^2}{3m}(a_2-a_0).
}
We then find the non-zero terms are $C_{\pm1,\pm1}^{\pm1,\pm1}=c_0+c_1$; $C_{00}^{00}=c_0$; $C_{\pm 1,\mp1}^{\pm1,\mp1}=c_0-c_1$; $C_{\pm1,0}^{\pm1,0}=c_0$; $C_{\pm1,0}^{0,\pm1}=c_1$; $C_{\pm1,\mp1}^{\mp1,\pm1}=0$, and finally $C^{\pm1,\mp1}_{0,0}=c_1$. 
\par
\emph{Hamiltonian terms}.--- Evaluating \eref{eomS2a}, for $m_{jk}=m/2$, we find
\EQ{\label{HmixEom}
\hbar d\phi_j(\rr)\Big|_{H_C}&=&-i{\cal P}_j \{L_C\phi_j(\rr)\}dt,
}
for $j=-1,0,1$, where the action of $L_C$ can be cast in standard form~\cite{Kawaguchi:2012bl}, as
\EQ{\label{Lcpm1}
L_C\phi_{\pm1}(\rr)&=&\left[\Heff_{\pm1}+c_0n(\rr)\pm c_1F_z(\rr)\right]\phi_{\pm1}(\rr)\nonumber\\
&&\;\;\;\;\;\;\;+\frac{c_1F_{\mp}(\rr)}{\sqrt{2}}\phi_{0}(\rr),\\
L_C\phi_{0}(\rr)&=&\left[\Heff_{0}+c_0n(\rr)\right]\phi_{0}(\rr)\nonumber\\
&&+\frac{c_1F_{+}(\rr)}{\sqrt{2}}\phi_{1}(\rr)+\frac{c_1F_{-}(\rr)}{\sqrt{2}}\phi_{-1}(\rr),
}
where $n(\rr)=|\phi_{-1}(\rr)|^2+|\phi_{0}(\rr)|^2+|\phi_{1}(\rr)|^2$, and the components of the spin density vector ${\bf F}\equiv (F_x,F_y,F_z)$ are
\EQ{\label{Fx}
F_x&=&\frac{1}{\sqrt{2}}\left[\phi_1^*\phi_0+\phi_0^*(\phi_1+\phi_{-1})+\phi^*_{-1}\phi_0\right],\\
F_y&=&\frac{i}{\sqrt{2}}\left[-\phi_1^*\phi_0+\phi_0^*(\phi_1-\phi_{-1})+\phi^*_{-1}\phi_0\right],\\
F_z&=&|\phi_1|^2-|\phi_{-1}|^2,
}
and $F_\pm\equiv F_x\pm iF_y=\sqrt{2}\left[\phi_{\pm1}^*\phi_0+\phi_{\mp1}\phi_0^*\right]$.
We thus find the PGPE for Hamiltonian $C$-field evolution of a spin-1 Bose gas.
\par
\emph{One-field dissipation.---} Evaluating the one-field terms, \eref{eomS2b}, noting that $\Gamma_\alp^{\nu\kap\sig}=\Gamma_\alp^{\nu\sig\kap}$, we then require $\Gamma_{\pm1}^{\pm1\pm1\pm1}=2(c_0+c_1)^2$; $\Gamma_0^{000}=2c_0^2$; $\Gamma_{\pm1}^{0,\pm1,0}=\Gamma_0^{\pm1,0,\pm1}=(c_0+c_1)^2/2$; $\Gamma_{\pm 1}^{\mp1,\pm1,\mp1}=(c_0-c_1)^2/2$, and $\Gamma_{\pm1}^{\mp1,0,0}=\Gamma_0^{0,\pm1,\mp1}=2c_1^2$. We then find the dissipation term
\EQ{\label{oneFieldTermS1}
\hbar d\phi_j(\rr)\Big|_{(1)}&=&{\cal P}_j\Big\{\gamma_{j}(\mu_j-L_C)\phi_j(\rr)dt+\hbar dW_j(\rr,t)\Big\},\;\;\;\;\;
}
where 
\EQ{\label{S1gam}
\gam_{\pm1}&=&\frac{4\pi}{\pi\lam_{dB}^2}\Big[2a_2^2\bar{G}_{\pm1,\pm1,\pm1}+a_2^2\bar{G}_{0,\pm1,0}\nonumber\\
&&+\left(\frac{2a_0+a_2}{3}\right)^2\bar{G}_{\mp1,\mp1,\pm1}+2\left(\frac{a_2-a_0}{3}\right)^2\bar{G}_{\mp1,0,0}\Big],\\
\gam_0&=&\frac{4\pi}{\pi\lam_{dB}^2}\Big[2\left(\frac{a_0+2a_2}{3}\right)^2\bar{G}_{0,0,0}+4\left(\frac{a_2-a_0}{3}\right)^2\bar{G}_{0,1,-1}\nonumber\\
&&+a_2^2\left(\bar{G}_{1,0,1}+\bar{G}_{-1,0,-1}\right)\Big].
}
In this form the effective cross-sections may be identified, and it is clear that distinguishable and indistinguishable collisions are correctly accounted for. In terms of these damping rates, the non-vanishing correlation function of the noise is
\EQ{\label{s1Noise}
dW_j^*(\rr,t)dW_k(\rr,t)&=&\frac{2\gamma_{j}k_BT}{\hbar}\delta_C^j(\rr,\rrp)\delta_{jk}dt.\;\;\;\;
}
The basic property of one-field damping is again evident in spin-1, namely that it is driven by additive, delta-correlated noise.
\par
%%--------------------------------------------------------------------
\begin{figure*}[!t]{
\begin{center} 
\includegraphics[width=1.7\columnwidth]{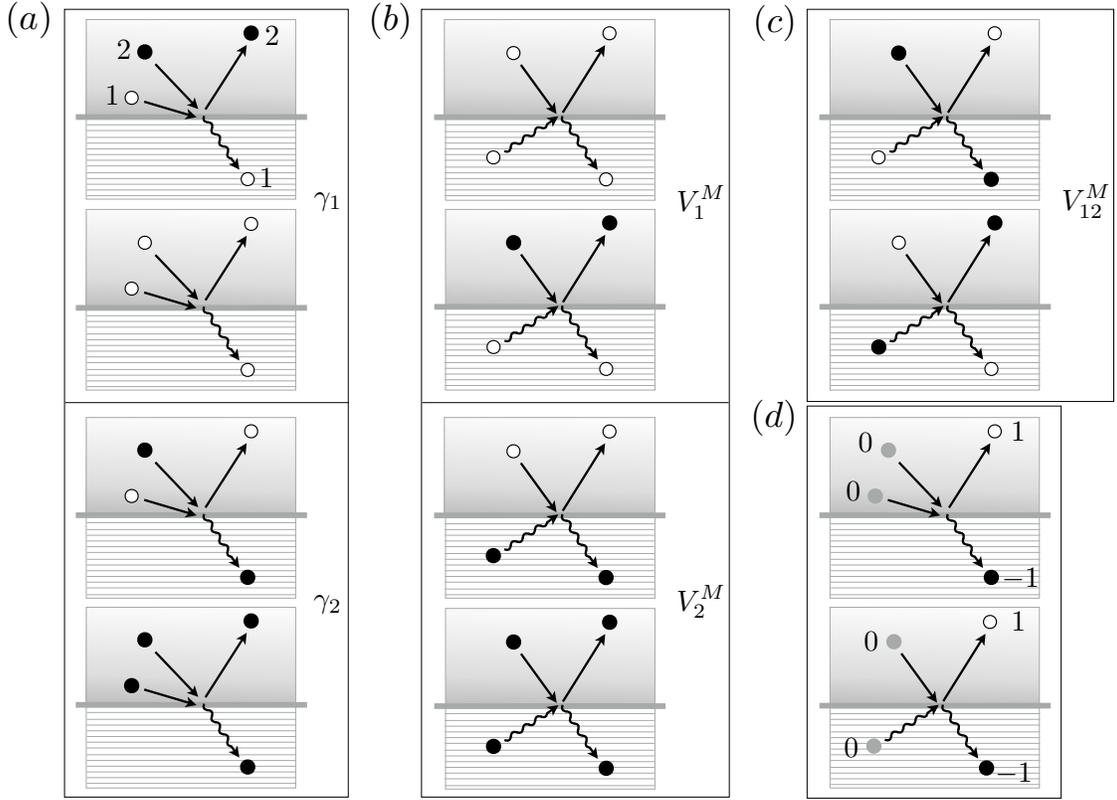}
\caption{Illustration of reservoir-interaction processes in two-component mixtures [(a), (b), (c)], and spin-1 condensates (d). (a) Processes contributing to one-field damping for the two-component mixture (the same processes occur in spinor systems for collisions between distinguishable particles). (b) Energy transfer processes arising from two-field damping. (c) Particle-transfer interactions arising from two-field damping. (d) Example dissipative processes in a spin-1 system stemming from spin-changing collisions in one- and two-field damping (respectively). Both processes generate particle exchange with the reservoir for the $-1$ component.
\label{fig2}}
\end{center}}
\end{figure*}
%%---------------------------------------------------------------
\emph{Two-field dissipation: energy damping.---} To evaluate the rate coefficients in \eref{eomS2c}, we require ${\rm V}_{\pm1,\pm1}^{\pm1}=2(c_0+c_1)^2$;  ${\rm V}_{\pm1,\pm1}^0=(c_0+c_1)^2$; ${\rm V}_{\pm1,\pm1}^{\mp1}=(c_0-c_1)^2$; ${\rm V}_{\pm1,0}^{\pm1}=(c_0+c_1)^2$; ${\rm V}_{\pm1,0}^0=2c_0(c_0+c_1)$; ${\rm V}_{\pm1,0}^{\mp1}=(c_0-c_1)(c_0+c_1)$; ${\rm V}_{1,-1}^{\pm1}=2(c_0-c_1)(c_0+c_1)$; ${\rm V}_{1,-1}^{0}=(c_0+c_1)^2$; ${\rm V}_{0,0}^{0}=4c_0^2$; ${\rm V}_{0,0}^{\pm1}=(c_0+c_1)^2$.
Evaluating the rates, we find
\EQ{\label{s1ED}
(S)\hbar d\phi_j(\rr)\Big|_{(2), {\rm V}}&=&{\cal P}_j\Big\{-iV_j^M(\rr)\phi_j(\rr)dt\nonumber\\
&&+i\hbar \phi_j(\rr)dU_j(\rr,t)\Big\},
}
where the energy-damping potential is
\EQ{\label{VMj}
V_j^M(\rr)&=&-\hbar\mint{\rrp} {\cal M}_{jk}(\rr-\rrp)\nabla\cdot \mathbf{J}_k(\rrp),
}
and the coefficients are
\EQ{\label{MjDefSpin}
{\cal M}_{jk}(\rr)&=&X_{jk}\int \frac{d^3\kk}{(2\pi)^3}\frac{e^{i\kk\cdot\rr}}{|\kk|},
}
with
\EQ{\label{xDef}
X_{jk}&\equiv&4\pi\left(\frac{m}{4\pi\hbar^2}\right)^2{\rm V}^q_{jk}\bar{\cal N}_{qq}.
}
The noise correlation is
\EQ{\label{dUjDef3}
dU_j(\rr,t)dU_k(\rrp,t)&=&\frac{2 k_BT}{\hbar}{\cal M}_{jk}(\rr-\rrp)dt.
}
We now see a new feature of spinor systems that is not evident in mixtures, namely, that the current-divergence of each component contributes to the energy damping potential, with weight \eref{xDef}. Evaluating these  weights, and noting that $X_{jk}=X_{kj}$, we find the potentials are determined by
\EQ{\label{chiEval}
X_{\pm1,\pm1}&=&4\pi\left[4a_2^2\bar{\cal N}_{\pm1,\pm1}+a_2^2\bar{\cal N}_{0,0}+\left(\frac{2a_0+a_2}{3}\right)^2\bar{\cal N}_{\mp1,\mp1}\right],\;\;\;\;\;\\
X_{\pm1,0}&=&4\pi\Bigg[2a_2^2\bar{\cal N}_{\pm 1,\pm1}+2a_2\left(\frac{a_0+2a_2}{3}\right)\bar{\cal N}_{0,0}\nonumber\\
&&+a_2\left(\frac{2a_0+a_2}{3}\right)\bar{\cal N}_{\mp1,\mp1}\Bigg],\\
X_{\pm1,\mp1}&=&4\pi\Bigg[2a_2\left(\frac{2a_0+a_2}{3}\right)\bar{\cal N}_{\pm 1,\pm1}+a_2^2\bar{\cal N}_{0,0}\nonumber\\
&&+2a_2\left(\frac{2a_0+a_2}{3}\right)\bar{\cal N}_{\mp1,\mp1}\Bigg],\\
X_{0,0}&=&4\pi\Bigg[a_2^2\bar{\cal N}_{1,1}+4\left(\frac{a_0+2a_2}{3}\right)^2\bar{\cal N}_{0,0}+a_2^2\bar{\cal N}_{-1,-1}\Bigg].
}

While we do not give a detailed treatment the particle-exchange terms given in \eref{eomS2d}, it is clear that there is a proliferation of new interaction processes for the spinor system due to the occurrence of spin-changing collisions. In Figure \ref{fig2} we give a schematic summary of the possible reservoir-interaction processes in the two-component mixture, and give examples of additional processes due to spin-changing interactions, described by \eref{eomS2d}, that can occur in the spin-1 system.
\section{Conclusions and outlook}\label{ConcOut}
In this paper we have generalised the SPGPE theory~\cite{Gardiner:2003bk} to systems of multi-component and spinor ultra-cold bosons subject to essentially arbitrary binary contact interactions. Our aim has been to present a complete treatment of the reservoir interaction problem for such systems, within a unified and tractable formalism, suitable for practical simulations of the experimental regime. The theory reveals an additional class of reservoir interaction processes (a generalisation of the process referred to as ``scattering"~\cite{Gardiner:2003bk}) whereby atoms from distinct $C$- and $I$-regions swap between regions, causing particle-exchange with the reservoir [see Fig.~\ref{fig1}~(b)]. This phenomenon is illustrated in the context of the two-component mixture. For spinor systems, the divergence of each component's current appears in the energy-damping potential, as shown explicitly for spin-1. This stronger inter-component coupling suggests that energy damping will play a more significant role in spinor systems, consistent with observations of energy damping in spin-1 experiments~\cite{Liu:2009en}. The theory presented here provides a general framework for Bose-Einstein condensates that can be used to find equations of motion for \emph{any} multi-component mixture, or spinor system of arbitrarily high spin. Full stochastic simulations of quenches and dissipative dynamics in spinor systems and mixtures are an important future aim. However we note that the purely damped equations, obtained by formally setting the noise in the SPGPE to zero, provide a qualitative description of the essential phenomenology of dissipative evolution~\cite{Tsubota2002,Penckwitt2002,Rooney:2010dp,Neely:2013ef,Rooney:2012gb,Rooney:2013ff,Neely:2013ef}.

\acknowledgments
We would like to acknowledge stimulating discussions with S. J. Rooney, M. C. Garrett, D. Baillie, and Y. Kawaguchi. We acknowledge financial supported from a Royal Society of New Zealand Rutherford Discovery Fellowship (ASB), and the Marsden Fund. 
\appendix 
\section{Dissipation rates for equal-mass collisions}\label{appA}
A somewhat simplified and high practical formulation can be obtained by assuming the reservoirs are approximately in equilibrium can thus be approximated as a Bose-Einstein distribution for some appropriate chemical potential and temperature. While seemingly a highly restrictive approach, this assumption does not greatly limit the dynamics of the $C$-region as the PGPE evolves both the condensate and a significant range of low energy non-condensate modes, and the $I$-region begins at a rather high energy, imposing a separation of the time scales characterising the $C$- and $I$-region evolution. Experimental quenches have also been successfully modelled using this approach~\cite{Weiler:2008eu}, and it may be reasonably expected to capture a great deal of the physics of the high temperature regime corresponding to the neighbourhood of the phase transition. A further advantage of this approach is that there is considerable simplification in the form of the rate functions, namely that $G(\rr)$ becomes spatially homogeneous over almost all of the $C$-region, and $M((\rr+\rrp)/2,\rr-\rrp,0)$ only depends on $\rr-\rrp$.
\par
In the next sections we evaluate the rate functions for the general case where each component can have a different $\mu_\sig,\ecut_\sig, V_\sig^{\rm eff}(\rr)$.
\subsection{One-field rate}
A significant simplification of the formalism is afforded by treating the $I$-regions as in thermal equilibrium. This may be a reasonable approximation for many systems, as the cutoff defining the $I$-region is set at quite high energy ($\sim 3\mu$), and thus there is a separation of timescales for the system evolution. 
\par
In the quasi-static regime, we expand the equilibrium Bose-Einstein distribution \eref{BEdef} to give
\EQ{\label{Gall}
G_{\nu\kap\sig}(\rr)&=&\frac{1}{2(2\pi)^5\hbar}\sum_{p=1}^\infty \sum_{q=1}^\infty\sum_{r=0}^\infty \nonumber\\
&&\times e^{ p\beta [\mu_\kap-\Veff_\kap(\rr)]+q\beta [\mu_\sig-\Veff_\sig(\rr)]
+r\beta [\mu_\nu-\Veff_\nu(\rr)]}\nonumber\\
&&\times \int_{I_\nu} d^3\kk_1\int_{I_\kap} d^3\kk_2\int_{I_\sig} d^3\kk_3\;\delta^{(3)}(\kk_1-\kk_2-\kk_3)\nonumber\\
&&\times \delta\left(\frac{\hbar^2}{2m}[\kk_1^2-\kk_2^2-\kk_3^2]-\Veff_{\nu\kap\sig}(\rr)\right)\nonumber\\
&& \times e^{ -r\beta\hbar^2\kk_1^2/2m-p\beta\hbar^2\kk_2^2/2m-q\beta\hbar^2\kk_3^2/2m},
}
where 
\EQ{\label{VeffDiff}
\Veff_{\nu\kap\sig}(\rr)&\equiv&\Veff_\kap(\rr)+\Veff_\sig(\rr)-\Veff_\nu(\rr).
}
Evaluating the momentum conservation delta-function gives $\kk_1\equiv \kk_2+\kk_3$, and 
\EQ{\label{Gdelta1}
G_{\nu\kap\sig}(\rr)&=&\frac{1}{2(2\pi)^5\hbar}\sum_{p=1}^\infty \sum_{q=1}^\infty\sum_{r=0}^\infty \nonumber\\
&&\times e^{ p\beta [\mu_\kap-\Veff_\kap(\rr)]+q\beta [\mu_\sig-\Veff_\sig(\rr)]+r\beta [\mu_\nu-\Veff_\nu(\rr)]}\nonumber\\
&&\times \int_{I_\kap} d^3\kk_2\int_{I_\sig} d^3\kk_3\; \delta\left(\frac{\hbar^2}{m}\kk_2\cdot\kk_3-\Veff_{\nu\kap\sig}(\rr)\right)\nonumber\\
&&\times e^{ -\beta\hbar^2/2m\left(r[\kk_2+\kk_3]^2+p\kk_2^2+q\kk_3^2\right)}\nonumber\\
&&\times \Theta\left[\frac{\hbar^2(\kk_2+\kk_3)^2}{2m}\geq \ecut_\nu-V_\nu(\rr)\right],
}
where the phase space restriction for the $\kk_1$ integral is expressed via the function $\Theta[a]\equiv 1$ when $a\geq 1$,and  $\Theta[a]\equiv0$ otherwise. Since the energy-conservation delta function imposes $\kk_2\cdot\kk_3\equiv m\Veff_{\nu\kap\sig}(\rr)/\hbar$, the $\Theta$-function condition can be written as
\EQ{\label{k23ineq}
\frac{\hbar^2}{2m}\left(\kk_2^2+\kk_3^2\right)\geq \ecut_\nu-\Veff_\kap(\rr)-\Veff_\sig(\rr).
}
Alternatively, using the $I_\kap$, $I_\sig$ phase-space restrictions gives
\EQ{\label{k23ineq2}
\frac{\hbar^2}{2m}\left(\kk_2^2+\kk_3^2\right)\geq \ecut_\kap+\ecut_\sig-\Veff_\kap(\rr)-\Veff_\sig(\rr),
}
and hence the condition \eref{k23ineq} always holds, provided
\EQ{\label{ecutCond}
\ecut_{\nu\kap\sig}\equiv \ecut_\kap+\ecut_\sig-\ecut_\nu\geq 0.
}
As the cutoffs are typically of a similar magnitude, it will usually be the case that this inequality will be satisfied, and hereafter it is presumed to hold. Choosing the $\kk_2$ and $\kk_3$ $z$-axes to coincide, the integral over their relative angle can be evaluated using the energy-conservation delta function to yield another $\Theta$ function. Changing variables to $s=(r+p)\beta\hbar^2k_2^2/2m$, $t=(r+q)\beta\hbar^2k_3^2/2m$ gives
\begin{widetext}
\EQ{\label{Gts}
G_{\nu\kap\sig}(\rr)&=&\frac{m^3}{(2\pi)^3\hbar^7\beta^2}\sum_{p=1}^\infty \sum_{q=1}^\infty\sum_{r=0}^\infty 
\frac{e^{ p\beta [\mu_\kap-\Veff_\kap(\rr)]+q\beta [\mu_\sig-\Veff_\sig(\rr)]+r\beta [\mu_\nu-\Veff_\kap(\rr)-\Veff_\sig(\rr)]}}{(r+p)(r+q)}\int_{s_{\rm min}(\rr)}^\infty ds\;e^{-s}\int_{t_{\rm min}(\rr,s)}^\infty dt \;e^{-t},
}
where
\EQ{\label{smin}
s_{\rm min}(\rr)&=&(r+p)\beta[\ecut_\kap-\Veff_\kap(\rr)],\\
\label{tmin}
t_{\rm min}(\rr,s)&=&(r+q)\beta {\rm max}\left\{\ecut_\sig-\Veff_\sig(\rr),\Veff_{\nu\kap\sig} (\rr)^2(r+p)\beta/4s\right\}.
}
\end{widetext}
A major simplification is obtained by noting that the integrals separate when $t_{\rm min}(\rr,s)\equiv (r+q)\beta[\ecut_\sig-\Veff_\sig(\rr)]$. This condition allows the rates to be evaluated in closed form, and also gives a result that is independent of $\rr$. Since $s\geq s_{\rm min}(\rr)$, this condition can be written as
\EQ{\label{VcondG}
\Veff_{\nu\kap\sig}(\rr)^2\leq4[\ecut_\sig-\Veff_\sig(\rr)][\ecut_\kap-\Veff_\kap(\rr)].
}
When \eref{VcondG} is satisfied, we arrive at the final result \eref{Gfinal}. 
\par
If all the trapping potentials and cutoffs are equal, the condition reduces to $\Veff(\rr)\leq 2\ecut/3$, as is known for the scalar SPGPE~\cite{Bradley:2008gq}. To estimate the region of validity for the scalar SPGPE, in a spherical harmonic trap with oscillator frequency $\omega$, the semiclassical turning point for the $C$-region is given by $\ecut=m\omega^2R_\epsilon^2/2$, and the condition holds for radii $|\rr|\leq \sqrt{2/3}R_\epsilon\approx 0.8R_\epsilon$, and hence the rate is position-independent over the bulk of the $C$-region. It can be shown that the rate reduces gradually near the edge of the $C$-region~\cite{Bradley:2008gq}. Thus for typical spinor systems where the trapping potentials will be very similar, if not identical, the position-independent expression \eref{Gfinal} describes the reservoir coupling rate for growth processes over the bulk of the $C$-region. 
\subsection{Two-field rate}
The two-field rate \eref{Mdef} may also be evaluated in the quasi-static reservoir approximation. We work with the fourier transform
\EQ{\label{Mft}
\bar{M}_{\kap\lam}(\rr,\kk)&\equiv&\mint{\vv}e^{-i\kk\cdot\vv}M_{\kap\lam}(\rr,\vv,0),
}
which, upon expansion of the Bose-Einstein distributions, gives
\EQ{\label{Mft2}
\bar{M}_{\kap\lam}(\rr,\kk)&=&\frac{1}{2(2\pi)^2\hbar}\sum_{p=1}^\infty\sum_{q=0}^\infty e^{p\beta[\mu_\kap-\Veff_\kap(\rr)]+q\beta[\mu_\lam-\Veff_\lam(\rr)]}\nonumber\\
&&\times\int_{I_\kap}d^3\kk_1\int_{I_\lam}d^3\kk_2\delta^{(3)}(\kk_1-\kk_2-\kk)\nonumber\\
&&\times \delta\left(\frac{\hbar^2}{2m}(\kk_1^2-\kk_2^2)+\Veff_\kap(\rr)-\Veff_\lam(\rr)\right).
}
Evaluating the $\kk_2$ integral gives $\kk_2\equiv\kk_1-\kk$. Making use of the energy $\delta$-function to simplify the new effective $\kk_2$ cutoff condition, the result may be expressed as
\EQ{\label{Mft3}
\bar{M}_{\kap\lam}(\rr,\kk)&=&\frac{1}{2(2\pi)^2\hbar}\sum_{p=1}^\infty\sum_{q=0}^\infty e^{p\beta[\mu_\kap-\Veff_\kap(\rr)]+q\beta[\mu_\lam-\Veff_\kap(\rr)]}\nonumber\\
&&\int_{I_\kap}d^3\kk_1\; e^{-\beta\hbar^2(p+q)\kk_1^2/2m}\Theta\left[\frac{\hbar^2\kk_1^2}{2m}\geq \ecut_\lam-\Veff_\kap(\rr)\right]\nonumber\\
&&\times \delta\left(\frac{\hbar^2}{2m}(\kk^2-2\kk\cdot\kk_1)+\Veff_{\lam}(\rr)-\Veff_{\kap}(\rr)\right).
} 
As in the previous section, in spherical coordinates the $z$-axis of $\kk_1$ can be taken along $\kk$, and the integral over the relative angle gives an additional cutoff condition:
\EQ{\label{angleTheta}
\int_0^\pi && \sin\theta\; d\theta\;\delta\left(\frac{\hbar^2}{2m}(k^2-2k k_1\cos\theta)+\Veff_{\lam}(\rr)-\Veff_{\kap}(\rr)\right)\nonumber\\
&&=\frac{m}{\hbar^2 k_1k}\Theta\left[\Bigg|\frac{\hbar^2\kk^2}{2m}+\Veff_\lam(\rr)-\Veff_\kap(\rr)\Bigg|\leq \frac{\hbar^2 k k_1}{m}\right].\;\;\;\;
}
The rate can then be written as
\EQ{\label{Mft4}
\bar{M}_{\kap\lam}(\rr,\kk)&=&\frac{m}{2(2\pi)\hbar^3|\kk|}\sum_{p=1}^\infty\sum_{q=0}^\infty e^{p\beta[\mu_\kap-\Veff_\kap(\rr)]+q\beta[\mu_\lam-\Veff_\kap(\rr)]}\nonumber\\
&&\times \int_{k_{\rm min}(\rr)}^\infty k_1\;dk_1\;e^{-\beta(p+q)\hbar^2\kk_1^2/2m},
}
where
\EQ{\label{kminM}
\frac{\hbar^2 k_{\rm min}(\rr)^2}{2m}&\equiv&{\rm max}\Big\{\ecut_\kap-\Veff_\kap(\rr),\;\ecut_\lam-\Veff_\kap(\rr),\nonumber\\
&&\frac{\hbar^2\kk^2}{8m}\left(1+\frac{2m[\Veff_{\lam}(\rr)-\Veff_{\kap}(\rr)]}{\hbar^2\kk^2}\right)^2\Big\}.\;\;\;\;
}
Assuming that the trapping potentials for each component are very similar, a straightforward application of the semi-classical argument given in Appendix A of Ref. \cite{Rooney:2012gb} shows that to a very good approximation the final term in Eq.~\eref{kminM} is always inferior to the larger of the first two terms. Hence, the rate becomes independent of $\rr$, taking the form
\EQ{
\bar{M}_{\kap\lam}(\rr,\k)=\bar{M}_{\kap\lam}(\k)&\equiv&\frac{{\cal N}_{\kap\lam}}{|\k|},
\label{Mgenfin}
}
where 
\EQ{\label{mcalDef}
{\cal N}_{\kap\lam}&\equiv&\left(\frac{ \pi m^2}{(2\pi)^2\hbar^5\beta}\right)\sum_{p=1}^\infty e^{p\beta(\mu_\kap-\bar{\epsilon}_{\kap\lam})}\Phi\left[e^{\beta(\mu_\lam-\bar{\epsilon}_{\kap\lam})},1,p \right],\;\;\;\;\;\;\;\;\;\\
\label{ecutBar}
\bar{\epsilon}_{\kap\lam}&\equiv&{\rm max}\left\{\ecut_\kap,\;\ecut_\lam\right\}.
}
We thus arrive at the quasi-static reservoir expression \eref{Mfinal}. Finally, to recover the scalar BEC case, where $\kap\equiv\lam$, it can be easily shown that
\EQ{\label{LerchId}
\sum_{p=1}^\infty z^p \Phi[z,1,p]=\frac{z}{1-z}.
}
%******************************************************************************
%\bibliographystyle{prsty}
%\bibliography{PapersRefsStandard}

\begin{thebibliography}{10}

\bibitem{Stenger:1998wt}
J. Stenger, S. Inouye, D. Stamper-Kurn, H. Miesner, A.~P. Chikkatur, and W.
  Ketterle, Nature {\bf 396},  345  (1998).

\bibitem{Lewandowski:2003hs}
H. Lewandowski, J. McGuirk, D. Harber, and E.~A. Cornell, Phys. Rev. Lett. {\bf
  91},  240404  (2003).

\bibitem{Schweikhard:2004gc}
V. Schweikhard, I. Coddington, P. Engels, S. Tung, and E.~A. Cornell, Phys.
  Rev. Lett. {\bf 93},  210403  (2004).

\bibitem{Saito:2006cm}
H. Saito, Y. Kawaguchi, and M. Ueda, Phys. Rev. Lett. {\bf 96},  4  (2006).

\bibitem{Liu:2009en}
Y. Liu, E. Gomez, S. Maxwell, L. Turner, E. Tiesinga, and P. Lett, Phys. Rev.
  Lett. {\bf 102},  225301  (2009).

\bibitem{Zhao:2013uk}
L. Zhao, J. Jiang, T. Tang, M. Webb, and Y. Liu,   (2013).

\bibitem{Endo:2011ip}
Y. Endo and T. Nikuni, J. Low Temp. Phys. {\bf 163},  92  (2011).

\bibitem{Kawaguchi:2012bl}
Y. Kawaguchi and M. Ueda, Physics Reports {\bf 520},  253  (2012).

\bibitem{StamperKurn:2013ku}
D.~M. Stamper-Kurn and M. Ueda, Rev. Mod. Phys. {\bf 85},  1191  (2013).

\bibitem{Beattie:2013ki}
S. Beattie, S. Moulder, R.~J. Fletcher, and Z. Hadzibabic, Phys. Rev. Lett.
  {\bf 110},  025301  (2013).

\bibitem{Matthews1999}
M.~R. Matthews, B.~P. Anderson, P.~C. Haljan, D.~S. Hall, C.~E. Wieman, and
  E.~A. Cornell, Phys. Rev. Lett. {\bf 83},  2498  (1999).

\bibitem{Myatt:1997ct}
C. Myatt, E. Burt, R. Ghrist, E. Cornell, and C. Wieman, Phys. Rev. Lett. {\bf
  78},  586  (1997).

\bibitem{Modugno:2002gz}
G. Modugno, M. Modugno, F. Riboli, G. Roati, and M. Inguscio, Phys. Rev. Lett.
  {\bf 89},  190404  (2002).

\bibitem{Simoni:2003hx}
A. Simoni, F. Ferlaino, G. Roati, G. Modugno, and M. Inguscio, Phys. Rev. Lett.
  {\bf 90},    (2003).

\bibitem{Papp:2008fg}
S.~B. Papp, J.~M. Pino, and C.~E. Wieman, Phys. Rev. Lett. {\bf 101},  040402
  (2008).

\bibitem{Pilch:2009cl}
K. Pilch, A. Lange, A. Prantner, G. Kerner, F. Ferlaino, H.~C. Nagerl, and R.
  Grimm, Phys. Rev. A {\bf 79},  042718  (2009).

\bibitem{McCarron:2011db}
D.~J. McCarron, H.~W. Cho, D.~L. Jenkin, M.~P. K{\"o}ppinger, and S.~L.
  Cornish, Phys. Rev. A {\bf 84},  011603  (2011).

\bibitem{Cho:2013fy}
H.-W. Cho, D. McCarron, M. K{\"o}ppinger, D. Jenkin, K. Butler, P. Julienne, C.
  Blackley, C. Le~Sueur, J. Hutson, and S. Cornish, Phys. Rev. A {\bf 87},
  010703  (2013).

\bibitem{Miesner:1999vz}
H.~J. Miesner, D. Stamper-Kurn, J. Stenger, S. Inouye, A.~P. Chikkatur, and W.
  Ketterle, Phys. Rev. Lett. {\bf 82},  2228  (1999).

\bibitem{Vengalattore:2010ti}
M. Vengalattore, J. Guzman, S.~R. Leslie, F. Serwane, and D.~M. Stamper-Kurn,
  Phys. Rev. A {\bf 81},  053612  (2010).

\bibitem{Guzman:2011kh}
J. Guzman, G.-B. Jo, A.~N. Wenz, K.~W. Murch, C.~K. Thomas, and D.~M.
  Stamper-Kurn, Phys. Rev. A {\bf 84},  063625  (2011).

\bibitem{Gardiner:2003bk}
C.~W. Gardiner and M.~J. Davis, J. Phys. B: At. Mol. Opt. Phys. {\bf 36},  4731
   (2003).

\bibitem{Bradley:2008gq}
A.~S. Bradley, C.~W. Gardiner, and M.~J. Davis, Phys. Rev. A {\bf 77},  033616
  (2008).

\bibitem{Bradley:2005jp}
A.~S. Bradley, P.~B. Blakie, and C.~W. Gardiner, J. Phys. B: At. Mol. Opt.
  Phys. {\bf 38},  4259  (2005).

\bibitem{Rooney:2010dp}
S.~J. Rooney, A.~S. Bradley, and P.~B. Blakie, Phys. Rev. A {\bf 81},  023630
  (2010).

\bibitem{Rooney:2012gb}
S.~J. Rooney, P.~B. Blakie, and A.~S. Bradley, Phys. Rev. A {\bf 86},  053634
  (2012).

\bibitem{Rooney:2014kc}
S.~J. Rooney, P.~B. Blakie, and A.~S. Bradley, Phys. Rev. E {\bf 89},  013302
  (2014).

\bibitem{Zaremba:1999iu}
E. Zaremba, T. Nikuni, and A. Griffin, J. Low Temp. Phys. {\bf 116},  277
  (1999).

\bibitem{Jackson:2001eg}
B. Jackson and E. Zaremba, Phys. Rev. Lett. {\bf 87},  100404  (2001).

\bibitem{Jackson:2002js}
B. Jackson and E. Zaremba, Phys. Rev. A {\bf 66},    (2002).

\bibitem{Jackson:2009jo}
B. Jackson, N.~P. Proukakis, C.~F. Barenghi, and E. Zaremba, Phys. Rev. A {\bf
  79},  053615  (2009).

\bibitem{Jackson:2007gy}
B. Jackson, N.~P. Proukakis, and C.~F. Barenghi, Phys. Rev. A {\bf 75},  051601
   (2007).

\bibitem{Proukakis:2008eo}
N.~P. Proukakis and B. Jackson, J. Phys. B: At. Mol. Opt. Phys. {\bf 41},
  203002  (2008).

\bibitem{Davis2001a}
M.~J. Davis, R.~J. Ballagh, and K. Burnett, J. Phys. B: At. Mol. Opt. Phys.
  {\bf 34},  4487  (2001).

\bibitem{Blakie05a}
P.~B. Blakie and M.~J. Davis, Phys. Rev. A {\bf 72},  063608  (2005).

\bibitem{Davis:2006ic}
M.~J. Davis and P.~B. Blakie, Phys. Rev. Lett. {\bf 96},  4  (2006).

\bibitem{Bezett09a}
A. Bezett and P.~B. Blakie, Phys. Rev. A {\bf 79},  023602  (2009).

\bibitem{Bezett09b}
A. Bezett and P.~B. Blakie, Phys. Rev. A {\bf 79},  033611  (2009).

\bibitem{Wright:2008ha}
T.~M. Wright, R.~J. Ballagh, A.~S. Bradley, P.~B. Blakie, and C.~W. Gardiner,
  Phys. Rev. A {\bf 78},  063601  (2008).

\bibitem{Wright:2009eh}
T.~M. Wright, A.~S. Bradley, and R.~J. Ballagh, Phys. Rev. A {\bf 80},
  (2009).

\bibitem{Wright:2011ey}
T.~M. Wright, N.~P. Proukakis, and M.~J. Davis, Phys. Rev. A {\bf 84},  023608
  (2011).

\bibitem{Wright:2010pj}
T.~M. Wright, A.~S. Bradley, and R.~J. Ballagh, Phys. Rev. A {\bf 81},
  (2010).

\bibitem{Wright:2012ud}
T.~M. Wright, M.~J. Davis, and N.~P. Proukakis,   (2012).

\bibitem{Wright:2010dd}
T.~M. Wright, P.~B. Blakie, and R.~J. Ballagh, Phys. Rev. A {\bf 82},  013621
  (2010).

\bibitem{Gawryluk:2007ig}
K. Gawryluk, M. Brewczyk, M. Gajda, and K. Rza{\.{z}}ewski, Phys. Rev. A {\bf
  76},  013616  (2007).

\bibitem{Pietila:2010ko}
V. Pietil{\"a}, T.~P. Simula, and M. M{\"o}tt{\"o}nen, Phys. Rev. A {\bf 81},
  033616  (2010).

\bibitem{Gardiner:2007gj}
S.~A. Gardiner and S.~A. Morgan, Phys. Rev. A {\bf 75},  043621  (2007).

\bibitem{Billam:2013kb}
T.~P. Billam, P. Mason, and S.~A. Gardiner, Phys. Rev. A {\bf 87},  033628
  (2013).

\bibitem{Mason:2014dd}
P. Mason and S.~A. Gardiner, Phys. Rev. A {\bf 89},  043617  (2014).

\bibitem{Weiler:2008eu}
C.~N. Weiler, T.~W. Neely, D.~R. Scherer, A.~S. Bradley, M.~J. Davis, and B.~P.
  Anderson, Nature {\bf 455},  948  (2008).

\bibitem{Davis:2012hq}
M.~J. Davis, P.~B. Blakie, A. van Amerongen, N. van Druten, and K.~V.
  Kheruntsyan, Phys. Rev. A {\bf 85},  031604(R)  (2012).

\bibitem{Rooney:2013ff}
S.~J. Rooney, T. Neely, B.~P. Anderson, and A.~S. Bradley, Phys. Rev. A {\bf
  88},  063620  (2013).

\bibitem{Blakie:2008is}
P.~B. Blakie, A.~S. Bradley, M.~J. Davis, R.~J. Ballagh, and C.~W. Gardiner,
  Adv. Phys. {\bf 57},  363  (2008).

\bibitem{Proukakis:2013dg}
N.~P. Proukakis, M.~J. Davis, and M. Szyma{\'{n}}ska, {\em {Quantum Gases}},
  Vol.~1 of {\em Cold Atoms} (Imperial College Press, London, 2013).

\bibitem{Rooney:2011fm}
S.~J. Rooney, P.~B. Blakie, B.~P. Anderson, and A.~S. Bradley, Phys. Rev. A
  {\bf 84},  023637  (2011).

\bibitem{Garrett:2013gk}
M.~C. Garrett, T.~M. Wright, and M.~J. Davis, Phys. Rev. A {\bf 87},  063611
  (2013).

\bibitem{Stoof:1999tz}
H.~T.~C. Stoof, J. Low Temp. Phys. {\bf 114},  11  (1999).

\bibitem{Bijlsma:2000vn}
M.~J. Bijlsma, E. Zaremba, and H. Stoof, Phys. Rev. A {\bf 62},  063609
  (2000).

\bibitem{Stoof:2001wk}
H.~T.~C. Stoof and M.~J. Bijlsma, J. Low Temp. Phys. {\bf 124},  431  (2001).

\bibitem{Duine2004}
R.~A. Duine, B.~W.~A. Leurs, and H.~T.~C. Stoof, Phys. Rev. A {\bf 69},  053623
   (2004).

\bibitem{Proukakis06a}
N.~P. Proukakis, J. Schmiedmayer, and H.~T.~C. Stoof, Phys. Rev. A {\bf 73},
  053603  (2006).

\bibitem{Cockburn09a}
S.~P. Cockburn and N.~P. Proukakis, Laser Phys {\bf 19},  558  (2009).

\bibitem{Cockburn10a}
S.~P. Cockburn, P.~G. Kevrekidis, N.~P. Proukakis, and D.~J. Frantzeskakis,
  Phys. Rev. Lett. {\bf 104},  174101  (2010).

\bibitem{Cockburn:2011kw}
S.~P. Cockburn, A. Negretti, N.~P. Proukakis, and C. Henkel, Phys. Rev. A {\bf
  83},    (2011).

\bibitem{Cockburn:2011fa}
S.~P. Cockburn, T. P~Horikis, P. G~Kevrekidis, N. P~Proukakis, and D.
  J~Frantzeskakis, Phys. Rev. A {\bf 84},  043640  (2011).

\bibitem{Cockburn:2012gc}
S.~P. Cockburn and N.~P. Proukakis, Phys. Rev. A {\bf 86},  033610  (2012).

\bibitem{Damski10a}
B. Damski and W.~H. Zurek, Phys. Rev. Lett. {\bf 104},  160404  (2010).

\bibitem{Su:2013dh}
S.-W. Su, S.-C. Gou, A.~S. Bradley, O. Fialko, and J. Brand, Phys. Rev. Lett.
  {\bf 110},  215302  (2013).

\bibitem{Su:2012jp}
S.~W. Su, I.~K. Liu, Y.~C. Tsai, W.~M. Liu, and S.~C. Gou, Phys. Rev. A {\bf
  86},  023601  (2012).

\bibitem{Ho:1996kj}
T.-L. Ho and V. Shenoy, Phys. Rev. Lett. {\bf 77},  3276  (1996).

\bibitem{QN}
C.~W. Gardiner and P. Zoller, {\em {Quantum Noise}}, 3rd  ed. (Springer-Verlag,
  Berlin Heidelberg, 2004).

\bibitem{Gardiner:2009wp}
C.~W. Gardiner, {\em {Handbook of Stochastic Methods}}, 4th ed.
  (Springer-Verlag, Berlin, 2009).

\bibitem{Hall:1998vx}
D.~S. Hall, M.~R. Matthews, J.~R. Ensher, C.~E. Wieman, and E.~A. Cornell,
  Phys. Rev. Lett. {\bf 81},  1539  (1998).

\bibitem{Sabbatini:2011gu}
J. Sabbatini, W. Zurek, and M.~J. Davis, Phys. Rev. Lett. {\bf 107},  230402
  (2011).

\bibitem{Tsubota2002}
M. Tsubota, K. Kasamatsu, and M. Ueda, Phys. Rev. A {\bf 65},  023603  (2002).

\bibitem{Penckwitt2002}
A.~A. Penckwitt, R.~J. Ballagh, and C.~W. Gardiner, Phys. Rev. Lett. {\bf 89},
  260402  (2002).

\bibitem{Neely:2013ef}
T.~W. Neely, A.~S. Bradley, E.~C. Samson, S.~J. Rooney, E.~M. Wright, K.~J.~H.
  Law, R. Carretero-Gonz{\'a}lez, P.~G. Kevrekidis, M.~J. Davis, and B.~P.
  Anderson, Phys. Rev. Lett. {\bf 111},  235301  (2013).

\end{thebibliography}

%******************************************************************************
\end{document}